\documentclass[12pt,a4paper]{article}

\usepackage[utf8]{inputenc}
\usepackage{amsmath, amssymb, amsthm, amsfonts}
\usepackage{graphicx} 
\usepackage{physics}
\usepackage{authblk}
\usepackage[svgnames, dvipsnames]{xcolor}
\usepackage{hyperref}
\usepackage{textcomp}
\usepackage{float}
\usepackage[margin=1in]{geometry}
\usepackage{array} 
\usepackage{changepage}
\usepackage{overpic}

\usepackage{tabularx}  
\usepackage{booktabs}  

\usepackage{orcidlink} 

\usepackage{caption}
\usepackage{subcaption}

\usepackage[
    natbib=true,
    style=numeric,
    maxnames=99,
    sorting=none
]{biblatex}
\title{The Vaccination Game on Networks}

\author[1,3]{Kausutua Tjikundi\orcidlink{0000-0001-9669-8789}}
\author[2]{Mark Broom\orcidlink{0000-0002-1698-5495}}

\affil[1]{\small \textit{ISI Foundation, Turin, Italy}}
\affil[2]{\small \textit{Department of Mathematics, City St. George's, University of London, London, UK}}
\affil[3]{\small \textit{Department of Computer Science, University of Turin, Turin, Italy}}

\date{} 

\begin{document}

\maketitle

\begin{abstract}
Vaccinations are an important tool in the prevention of disease. Vaccinations are generally voluntary for each member of a population and vaccination decisions are influenced by individual risk perceptions and contact structures within populations. In this study, we model vaccination uptake as an evolutionary game where individuals weigh perceived morbidity risks from both vaccination and infection. We incorporate epidemiological dynamics using an SIR model structured on networks, allowing us to determine the evolutionarily stable vaccination level for any given network topology. Our analysis shows that vaccination coverage varies across networks depending on their structure and the relative cost of vaccination (the ratio of vaccine morbidity risk to infection morbidity risk). As this cost increases, vaccination uptake decreases universally, leading to a dominant non-vaccinator strategy when the cost is high. We find that networks with low to moderate degree variability have a relatively low evolutionarily stable vaccination level when this cost is low, as in such populations lower vaccination levels are necessary to achieve equivalent levels of disease prevalence to more heterogeneous networks with high degree variability, which thus show higher vaccination levels at lower relative costs. However, for heterogeneous networks, vaccination levels decline faster as costs rise, eventually falling below the level for the more homogeneous networks. Our findings align with previous studies on vaccination thresholds in structured populations and highlight how network heterogeneity influences vaccination dynamics.

\end{abstract}


\section{Introduction}

Modern scientific and technological progress has made it possible to develop safe and effective vaccines for numerous prevalent infectious diseases, including smallpox, measles, pertussis, influenza, and chickenpox, and the widespread use of these vaccines is estimated to save millions of lives every year \cite{bonanni1999demographic}. While significant strides have been made, considerable challenges remain, mainly due to the difficulties encountered in administering and disseminating vaccines in the world's most impoverished and conflict-ridden regions \cite{cousins2015syrian}, but also because in some parts of the world, there is still a significant amount of vaccine refusal or hesitancy among individuals, often due to baseless worries about severe side effects such as autism. 
Even though vaccination is one of the most significant preventative health measures that have ever been implemented globally to stop the spread of infectious diseases that have severely harmed many human societies in the recent past \cite{wang2017vaccination, anderson1991infectious}, it is always voluntary and hence individuals have an opportunity to say no. The study \cite{mckee2016exploring} identified four major reasons that lead some individuals to not get vaccinated: religious concerns, philosophical concerns, safety concerns due to side effects, and insufficient knowledge. Recently a qualitative analysis of social media posts in Germany revealed six main categories of reasons for rejecting COVID-19 vaccination: low perceived benefit of vaccination, low perceived risk of contracting COVID-19, health concerns, lack of information, systemic mistrust, and spiritual or religious reasons \cite{fieselmann2022reasons}. 

The decision to accept vaccination is thus often influenced by individuals' perceptions of the potential risks associated with it \cite{fine1986individual, chapman1999predictors, chapman2006emotions, basu2008integrating}.
The decision to get vaccinated can be thought of as analogous to the Prisoner's dilemma, a classic game theory scenario where individuals have two choices: cooperate (get vaccinated) or defect (not) \cite{bauch2004vaccination, broom2022game}, where defecting (non-vaccinating) individuals act as free riders, benefiting from the behaviour of the cooperating (vaccinating) individuals. We discuss this aspect for the vaccination game below.

Most of the previous game theoretical studies \cite{bauch2003group, bauch2004vaccination, bauch2005imitation, reluga2006evolving, galvani2007long, vardavas2007can, han2014can} assumed a homogeneous population, which offers a simplified, theoretical framework for understanding transmission dynamics but can be unrealistic. This assumption neglects the important role of social structure in disease spread and individual behaviour. In real-world scenarios, individuals interact within networks, and the structures of the networks influence the vaccination behaviour.  For instance, networks with shorter average path lengths are found to have higher vaccination rates because diseases spread faster on these types of networks, hence increasing the perceived risk of infection and making vaccination a more appealing strategy \cite{tanimoto2015pandemic}. The network structure plays an important role in shaping how infectious diseases spread through a human population \cite{keeling2005networks, danon2011networks, wang2016statistical, bellingeri2024critical} and can also influence or change individuals’ potential motivation to get vaccinated \cite{perisic2009simulation} and there is a growing effort to use network-based frameworks to understand the interplay between epidemic spread and human behaviour relating to vaccination \cite{zhang2010hub, zhang2012rational, zhang2014effects, shi2017voluntary}. Network models that account for the heterogeneity of connections among people are beneficial for studying how diseases propagate. By applying such models to the vaccination game, we can capture the individual-level variation in the transmission dynamics of infectious diseases \cite{kabir2020impact}. 

The issue at hand involves individuals making decisions, taking into account their knowledge of vaccination and the vaccines. While these vaccines offer a high level of protection against infection, they may also be associated with side effects. As a result, individuals are often faced with the decision of weighing the benefits of protection against the potential negative consequences of receiving these vaccinations. Additionally, it is worth noting that once an individual is vaccinated, they are not only protected but also offering protection to others and this relates to an important public good concept referred to as herd immunity \cite{fu2011imitation, morison2025public}. Thus, vaccination generates a public good of herd immunity and once established, those that are still not vaccinated enjoy the benefit of protection without paying the cost \cite{tanimoto2015pandemic}. However, this scenario creates a temptation to free-ride (a situation where individuals may rationally decide to forgo vaccination and rely on others to get vaccinated). A population with a sufficiently high vaccine coverage can eradicate a disease without vaccinating everyone. As coverage increases, individuals have a greater incentive not to vaccinate because non-vaccinators can reap the benefits of herd immunity without the perceived possibility of experiencing adverse effects or complications from the vaccine \cite{bauch2005imitation, tanimoto2015pandemic}. We note that in most models, including the original work of \cite{bauch2004vaccination}, this process means that the level of vaccination can never be high enough to reach the level required to eradicate the disease.

In this study, we model vaccination as a game where payoffs are based on perceived morbidity risks from both the vaccine and the infection. We incorporate epidemiological dynamics using the SIR model (Susceptible, Infectious, Recovered) with a network structure. The goal is that, given any network structure (represented by an adjacency matrix $A_{ij}$), we can to determine the appropriate vaccination level that would solve our game (find the evolutionarily stable vaccination level), based on the relative cost of vaccination (the ratio of the vaccine morbidity risk to the infection morbidity risk). We can also identify a threshold perceived risk above which it is optimal for individuals not to vaccinate.

\section{The vaccination model}

Following \cite{bauch2004vaccination}, suppose an individual's probability of choosing to get vaccinated is denoted by $P$, and the proportion of the population that has already been vaccinated is represented by $p$. Vaccination carries a morbidity risk, denoted by $r_v >0$, while contracting the infection without vaccination carries a morbidity risk of $r_i > 0$. The probability of an unvaccinated individual becoming infected (when the vaccine coverage is $p$) is denoted by $\pi_{p}$. It is assumed that the vaccine is completely effective, meaning that there is zero probability of a vaccinated individual contracting the disease. The payoff for a vaccinated individual is simply minus the cost it pays for getting vaccinated $-r_v$, and the payoff for an unvaccinated individual is minus the cost of catching the disease multiplied by the probability of catching it, $-r_i\pi_{p}$.

The expected payoff for vaccinating with probability $P$ is thus given as follows: 
\begin{equation}
    E(P,p) = P(-r_{v}) + (1-P)(-r_{i}\pi_{p}).
\end{equation}
Dividing the payoff function above by the constant $r_{i}$ (note that the game is unchanged if we scale the payoff function by a constant) and denoting $r = r_{v}/r_{i}$ as the relative cost of vaccination we obtain the following standardised payoff,
\begin{equation}
E(P,p) = -rP - \pi_{p}(1-P) = - \pi_{p} + P(\pi_{p} - r).
\end{equation}

The difference 
\begin{equation}\label{eq:incentive}
E[1,p] - E[0,p] = \pi_{p} - r   
\end{equation}
is the {\it incentive function} for vaccination. It represents the difference in payoffs between vaccinated and unvaccinated individuals. If this value is positive, it means that the benefits of vaccination (reduced risk of infection) outweigh the costs, providing an incentive for individuals to get vaccinated.


It was shown in Bauch and Earn \cite{bauch2004vaccination} that setting equation \eqref{eq:incentive} equal to zero gives a unique solution, which is the evolutionarily stable solution to the vaccination game.

\section{The SIR model dynamics }
As in \cite{bauch2004vaccination}, we consider a standard three-compartment model in which individuals
are either susceptible to the disease $(S)$, infectious $(I)$, or recovered to a state of lifelong immunity $(R)$.
\begin{align}
\frac{dS}{dt} &= \mu(1 - p) - \beta SI - \mu S, \label{eq:dSdt} \\
\frac{dI}{dt} &= \beta SI - \gamma I - \mu I, \label{eq:dIdt} \\
\frac{dR}{dt} &= \mu p + \gamma I - \mu R. \label{eq:dRdt}
\end{align}

Here, $\mu$ is the mean birth and death rate (the birth and death rates are equal to ensure a fixed population size), $\beta$ is the mean transmission rate, $1/ \gamma$ is the mean infectious period, and $p$ is the vaccine uptake level (for simplicity we assume that individuals are never infected before being vaccinated). Once a dynamical steady state is reached, the vaccine coverage level in the population will equal the uptake level.

In the model of Bauch and Earn \cite{bauch2004vaccination} the probability of infection for the above model was shown to be
\begin{equation}\label{eq:equationforpip}
\pi_{p}=1-\frac{\gamma+\mu}{\beta (1-p)}.
\end{equation}

\subsection{Incorporating the network structure}

To incorporate the adjacency matrix into the model, we extend the model to represent a network of interacting individuals. We note here that $\beta$ represents the rate at which the disease is passed between any pair of individuals, so for instance to compare results from a regular graph with degree $d$ to the original model we would need to scale $\beta$ by dividing by $d$.
Considering a network with \(N\) nodes, we modify equations \eqref{eq:dSdt}, \eqref{eq:dIdt}, and \eqref{eq:dRdt} to obtain the following equations:
\begin{align}
\frac{dS_i}{dt} &= \mu(1 - p_i) - \beta S_i\left(\sum_{j=1}^{N} A_{ij}X_j\right) - \mu S_i, \label{eq:dSi_dt} \\
\frac{dX_i}{dt} &= \beta S_i\left(\sum_{j=1}^{N} A_{ij}X_j\right) - \gamma X_i - \mu X_i, \label{eq:dXi_dt} \\
\frac{dR_i}{dt} &= \mu p_i + \gamma X_i - \mu R_i. \label{eq:dRi_dt}
\end{align}

In these equations: \(S_i\) represents the probability that node \(i\) is susceptible, \(X_i\) (instead of "$I_i$") is the probability that node \(i\) is infectious, \(R_i\) denotes the probability that node \(i\) is recovered/immune, \(p_i\) represents the probability that the individual at node $i$ is vaccinated and \(A_{ij}\) is the adjacency matrix element representing the connection between nodes \(i\) and \(j\). Here $A_{ij}=1$ if individuals are connected and so can potentially infect each other, and $A_{ij}=0$ otherwise. Equation \eqref{eq:dSi_dt} controls the rate of change of the susceptible probability $(S_i)$, equation \eqref{eq:dXi_dt} represents the rate of change of the infectious probability $(X_i)$, and equation \eqref{eq:dRi_dt} represents the rate of change of the recovered/immune probability $(R_i)$ for node \(i\) at time $t$.  These equations thus extend the vaccination game model of \cite{bauch2004vaccination} to account for interactions between nodes in a network, where the transmission of the disease depends on the connectivity represented by the adjacency matrix.

\subsection{The steady state solution}
To find the steady-state solution of the model, we set the above derivatives with respect to time equal to zero and solve the resulting system of equations for the steady-state values of \(S_i\), \(X_i\), and \(R_i\). We obtain:
\begin{align}
0 &= \mu(1 - p_i) - \beta S_i\left(\sum_{j=1}^{N} A_{ij}X_j\right) - \mu S_i, \label{eq:Si_eq} \\
0 &= \beta S_i\left(\sum_{j=1}^{N} A_{ij}X_j\right) - \gamma X_i - \mu X_i, \label{eq:Xi_eq} \\
0 &= \mu p_i + \gamma X_i - \mu R_i. \label{eq:Ri_eq}
\end{align}

Now, we have a system of equations to solve for the steady-state values of \(S_i\), \(X_i\), and \(R_i\). 
We need to find these steady-state solutions for a given set of parameter values and network structure. In practice, we need to specify the values of \(\mu\), \(\beta\), \(\gamma\), the adjacency matrix \(A_{ij}\), and the vaccination probabilities \(p_i\) to find these steady-state proportions. Solving for \(S_i\) in equation \eqref{eq:Si_eq} at steady state and rearranging, we obtain:

\begin{equation}
S_i = \frac{\mu(1 - p_i)}{\mu + \beta \left(\sum_{j=1}^{N} A_{ij}X_j\right)}.
\label{eq:Si_only}
\end{equation}
Solving for \(X_i\) using equation \eqref{eq:Xi_eq} at steady state, we get:
\begin{equation}
X_i = \frac{\beta}{\gamma + \mu} S_i \left(\sum_{j=1}^{N} A_{ij}X_j\right).
\label{eq:Xi_only}
\end{equation}

Substituting \eqref{eq:Si_only} into \eqref{eq:Xi_only} yields:
\begin{align}
X_i &= \frac{\beta}{\gamma + \mu} 
\frac{\mu(1 - p_i) \sum_{j=1}^{N} A_{ij}X_j}{\mu + \beta \sum_{j=1}^{N} A_{ij}X_j}. 
\label{eq:Xi_final}
\end{align}
Since $S_i+X_i+R_i=1$, the solutions to the above equations will automatically give the value of $R_i$.

Now we take equation \eqref{eq:Xi_final} for $X_{i}$ and multiply through by $\mu + \beta \sum_{j=1}^{N} A_{ij}X_j$ and then divide everything by $\mu$ to obtain 
\begin{equation}
X_i + \frac{\beta}{\mu} X_i\sum_{j=1}^{N} A_{ij}X_j = \frac{\beta}{\gamma + \mu} (1 - p_i) \sum_{j=1}^{N} A_{ij}X_j. \label{eq:infec_second}
\end{equation}
Setting $b = \frac{\beta}{\mu}$ and $c = \frac{\beta}{\gamma + \mu}$ yields:
\begin{equation}
X_i + bX_i\sum_{j=1}^{N} A_{ij}X_j = c(1 - p_i) \sum_{j=1}^{N} A_{ij}X_j.
\label{eq:infec_final}
\end{equation}

Equation \eqref{eq:infec_final} gives a general equation based upon potentially different vaccination probabilities at each node. In the analysis that follows, we assume that this vaccination probability is constant over all nodes, i.e. that $p_{i}=p$ for all values of $i$, representing the  overall population vaccination coverage. Thus, whilst individuals are able to make decisions based upon the overall perceived risk, they cannot calculate their own individual risk based upon their individual connections. Thus, equation \eqref{eq:infec_final} becomes
\begin{equation}
X_i\left( 1+  b\sum_{j=1}^{N} A_{ij}X_j \right) = c(1 - p) \sum_{j=1}^{N} A_{ij}X_j.
\label{eq:infec_finalpsame}
\end{equation}
This overall perceived risk is then simply given by the probability of a randomly selected individual in the population to catch the disease given that they are unvaccinated, which as before, we denote by $\pi_{p}$.

The probability that a given individual catches the disease is the ratio of the rate that they catch the disease and the sum of this rate and its rate of death, i.e.
\begin{equation}\label{eq:requation}
\frac{\beta \sum_j A_{ij} X_j}{\mu + \beta \sum_j A_{ij} X_j}=\frac{b \sum_j A_{ij} X_j}{1 + b \sum_j A_{ij} X_j}.
\end{equation}

Combining this with the fact that the evolutionarily stable vaccination level equates the costs from the two alternative actions, solving $\pi_{p}=r$, 
leads to the following equation involving $r$ and the average probability of catching the disease:
\begin{equation}\label{eq:requation}
r = \pi_p = \frac{1}{N} \sum_i \frac{b \sum_j A_{ij} X_j}{1 + b \sum_j A_{ij} X_j}.
\end{equation}

\subsection{A regular graph with degree d}
Here we derive an analytical solution for $p$ from \eqref{eq:infec_final} using a regular graph, where each vertex has the same number of neighbours (the degree of the vertex). In a regular graph with degree $d$, each row of the adjacency matrix $A_{ij}$ will sum to $d$, as each node has $d$ connections. We assume that $X_i = x$ and $p_i = p$. Thus, the sum $\sum_{j=1}^{N} A_{ij}X_j$ simplifies to $dx$ and equation \eqref{eq:infec_final} reduces to:
\begin{align}
x + bx^2 d &= c(1 - p)dx \Rightarrow \label{eq:first} \\
x &= \frac{cd(1 - p) - 1}{bd}. \label{eq:third}
\end{align}

Equation \eqref{eq:requation} then yields:
\begin{equation}
r = \frac{\beta dx}{\mu + \beta dx} 
= \frac{bdx}{1 + bdx}. \label{eq:r_early}
\end{equation}
Substituting for $x$ in equation \eqref{eq:third} using equation \eqref{eq:third} gives
\begin{equation}
r = 1 - \frac{1}{cd(1 - p)}
\label{eq:r_complete}
\end{equation}
which implies that 
\begin{equation}
p = 1 - \frac{1}{cd(1 - r)}.
\label{eq:p_complete}
\end{equation}
The above solution of course holds only for $p$ values between 0 and 1, and it follows straightforwardly from the calculations that if the solution to equation \eqref{eq:p_complete} is less than 0 the optimal $p$ value is simply equal to 0.

\subsection{Bipartite graphs} 

Consider a complete bipartite graph. Here, vertices are divided into two sets, with $d_{2}$ and $d_{1}$ members, respectively. Each vertex is connected to all vertices from the other set, but none from their own, so that all members of the first set have degree $d_1$ and all members of the second set have degree $d_2$. Assuming that all individuals have the same vaccination probability $p$, all members of each set will have the same probability of infection; denote this by $x$ for the first set and $y$ for the second. Using equation \eqref{eq:infec_final} we thus obtain:
\begin{equation}\label{eq:bipxeq}
x(1+bd_{1}y)=c(1-p)d_{1}y    
\end{equation}
and 
\begin{equation}\label{eq:bipyeq}
y(1+bd_{2}x)=c(1-p)d_{2}x.
\end{equation}
Solving equations (\ref{eq:bipxeq}) and (\ref{eq:bipyeq}) simultaneously, we obtain
\begin{equation}\label{eq:xyform}
x=\frac{c^{2}(1-p)^{2}d_{1}d_{2}-1}{bd_{2}(1+c(1-p)d_{1})}, \hspace{0.1cm} y=\frac{c^{2}(1-p)^{2}d_{1}d_{2}-1}{bd_{1}(1+c(1-p)d_{2})}.
\end{equation}    
Further, using equation \ref{eq:requation} we obtain
\begin{equation}\label{eq:biprxy}
r=\frac{1}{d_{1}+d_{2}} \left(d_{2}\frac{bd_{1}y}{1+bd_{1}y}+d_{1}\frac{bd_{2}x}{1+bd_{2}x} \right).    
\end{equation}
Substituting equation (\ref{eq:xyform}) into equation (\ref{eq:biprxy}), we obtain the following formula for $r$ as a function of $p$.
\begin{equation}\label{eq:r_bipartite}
r=\frac{c^{2}d_{1}d_{2}(1-p)^{2}-1}{(d_{1}+d_{2})c(1-p)} \left( \frac{1}{(1+cd_{1}(1-p)}+ \frac{1}{(1+cd_{2}(1-p)}\right)
\end{equation}

We are, of course, interested in finding $p$ as a function of $r$ rather than the other way around, but there is no simple closed-form solution to this, and it is thus convenient to represent the condition in the above form.

\begin{figure}[h]
    \vspace{0.2cm}
    \begin{subfigure}{0.32\textwidth}
       \includegraphics[width=\textwidth]{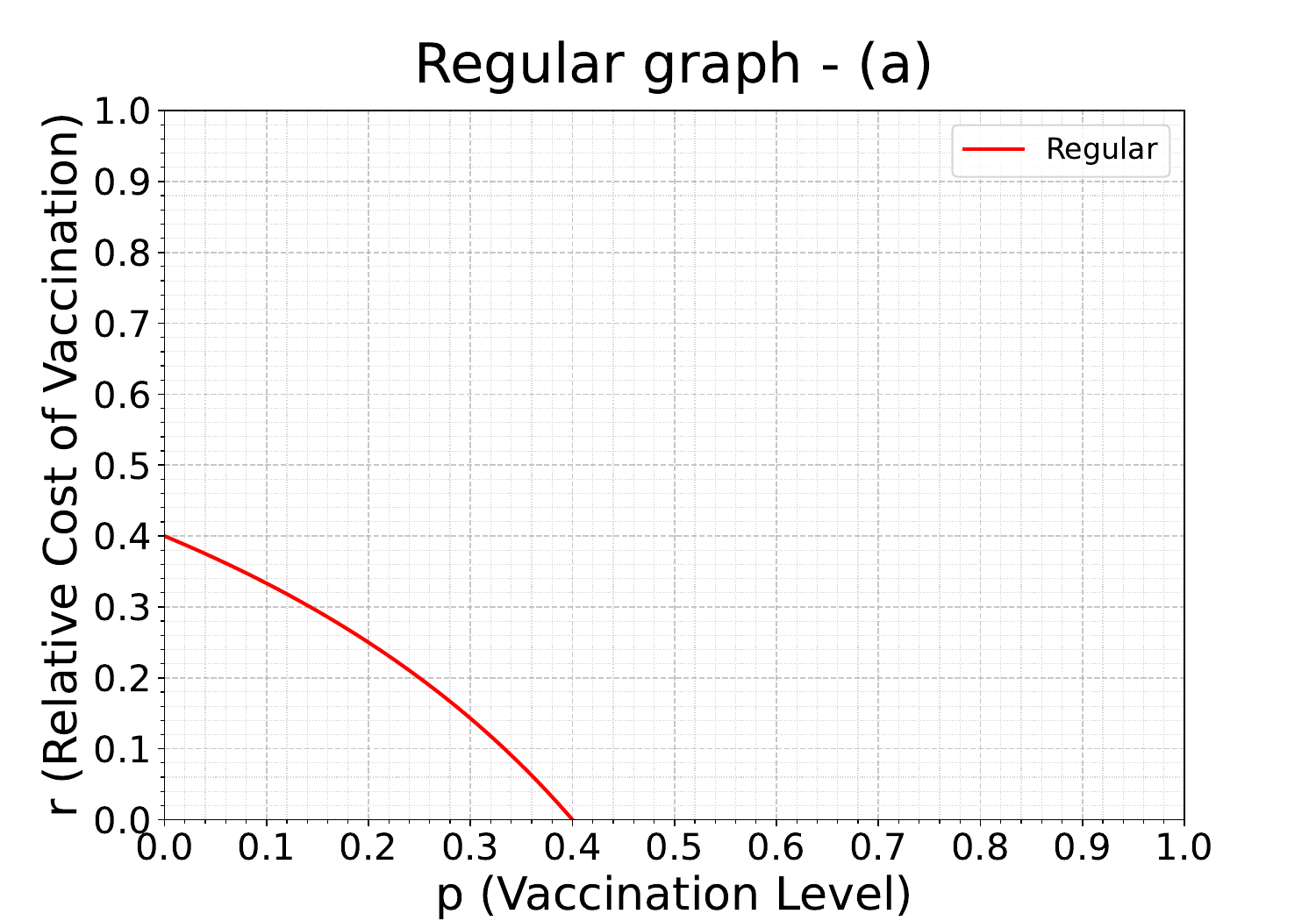}
    \end{subfigure}\hfill
    \begin{subfigure}{0.32\textwidth}
       \includegraphics[width=\textwidth]{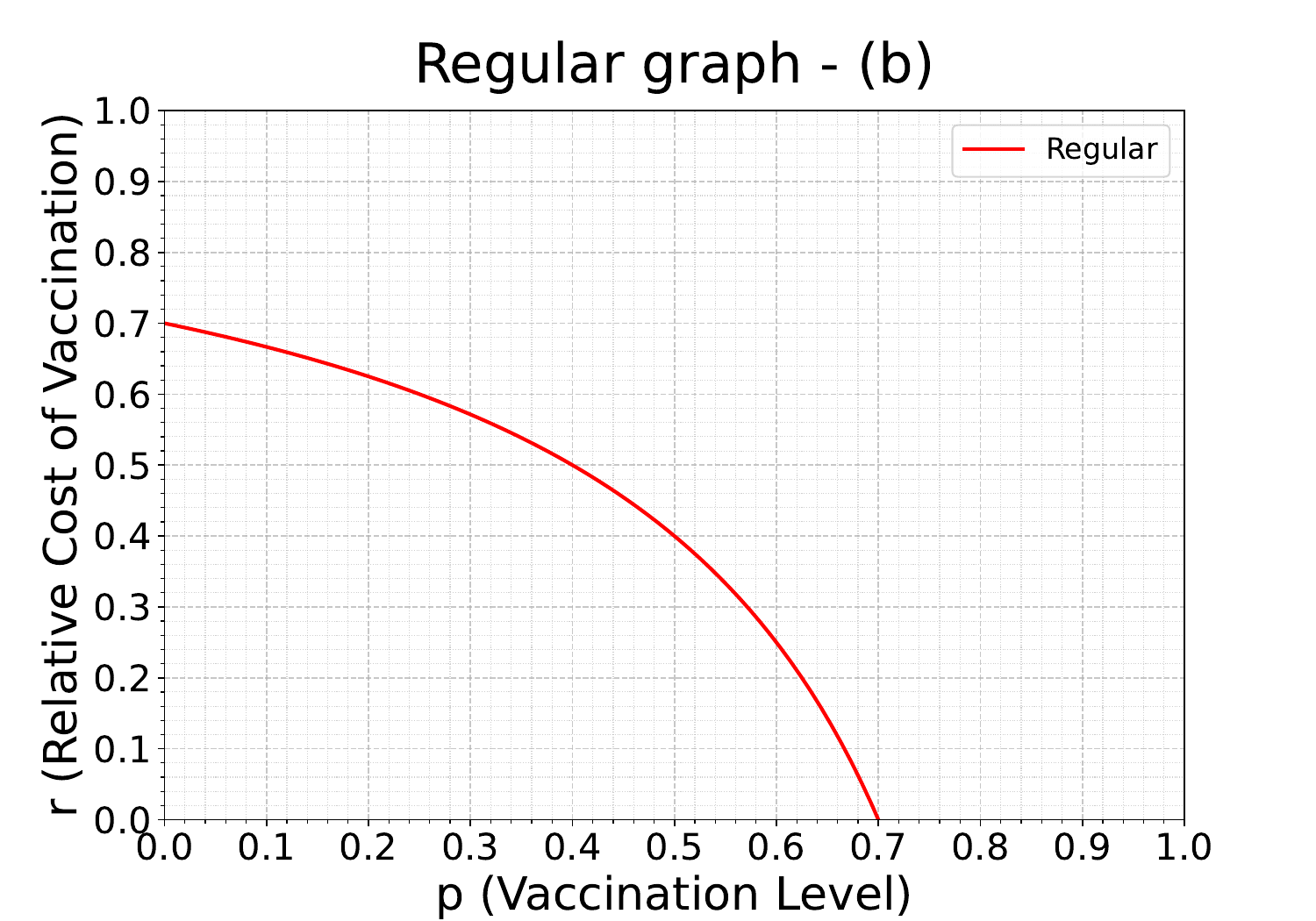}
    \end{subfigure}\hfill
    \begin{subfigure}{0.32\textwidth}
       \includegraphics[width=\textwidth]{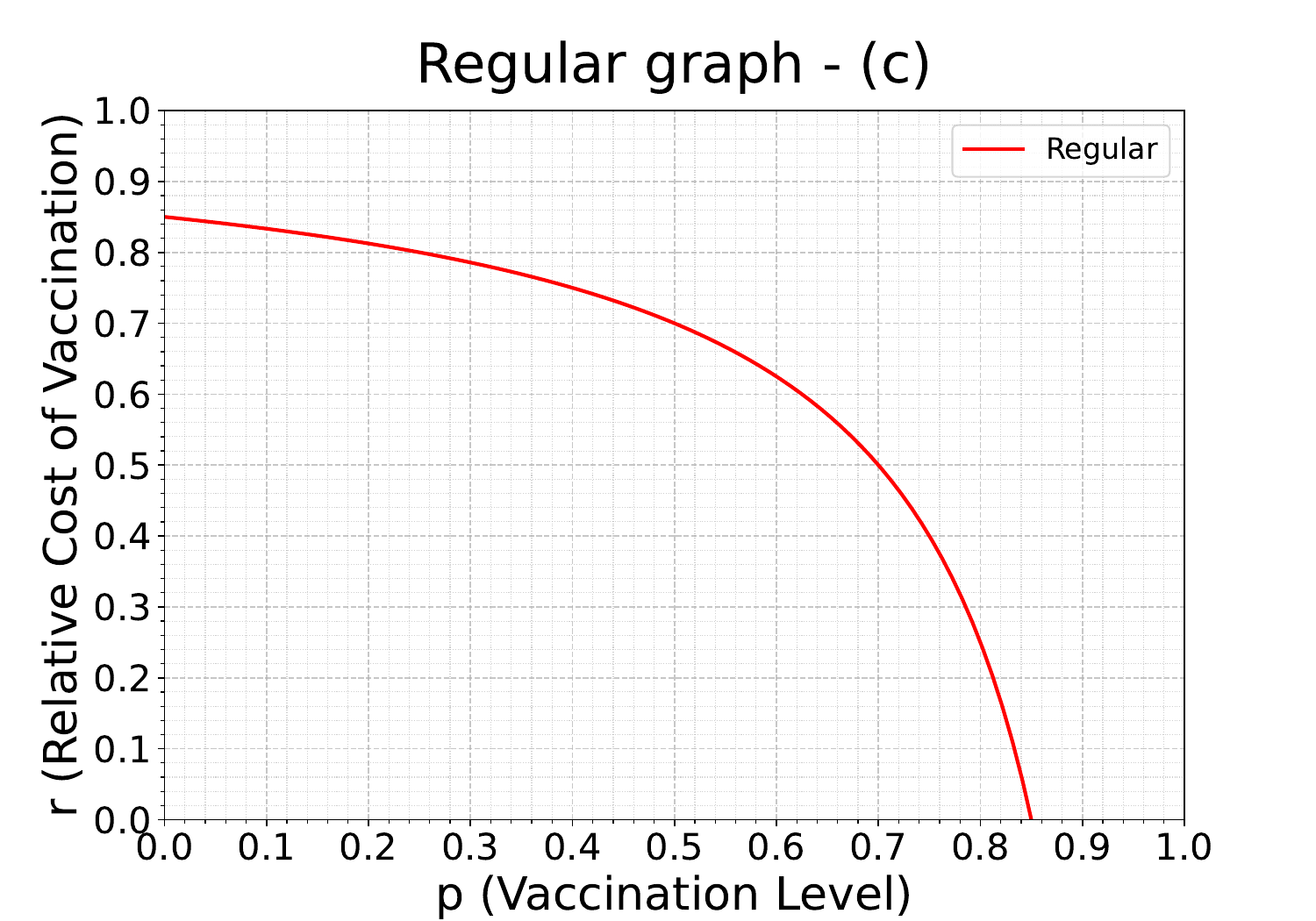}
    \end{subfigure}

    \bigskip 
    \begin{subfigure}{0.32\textwidth}
       \includegraphics[width=\textwidth]{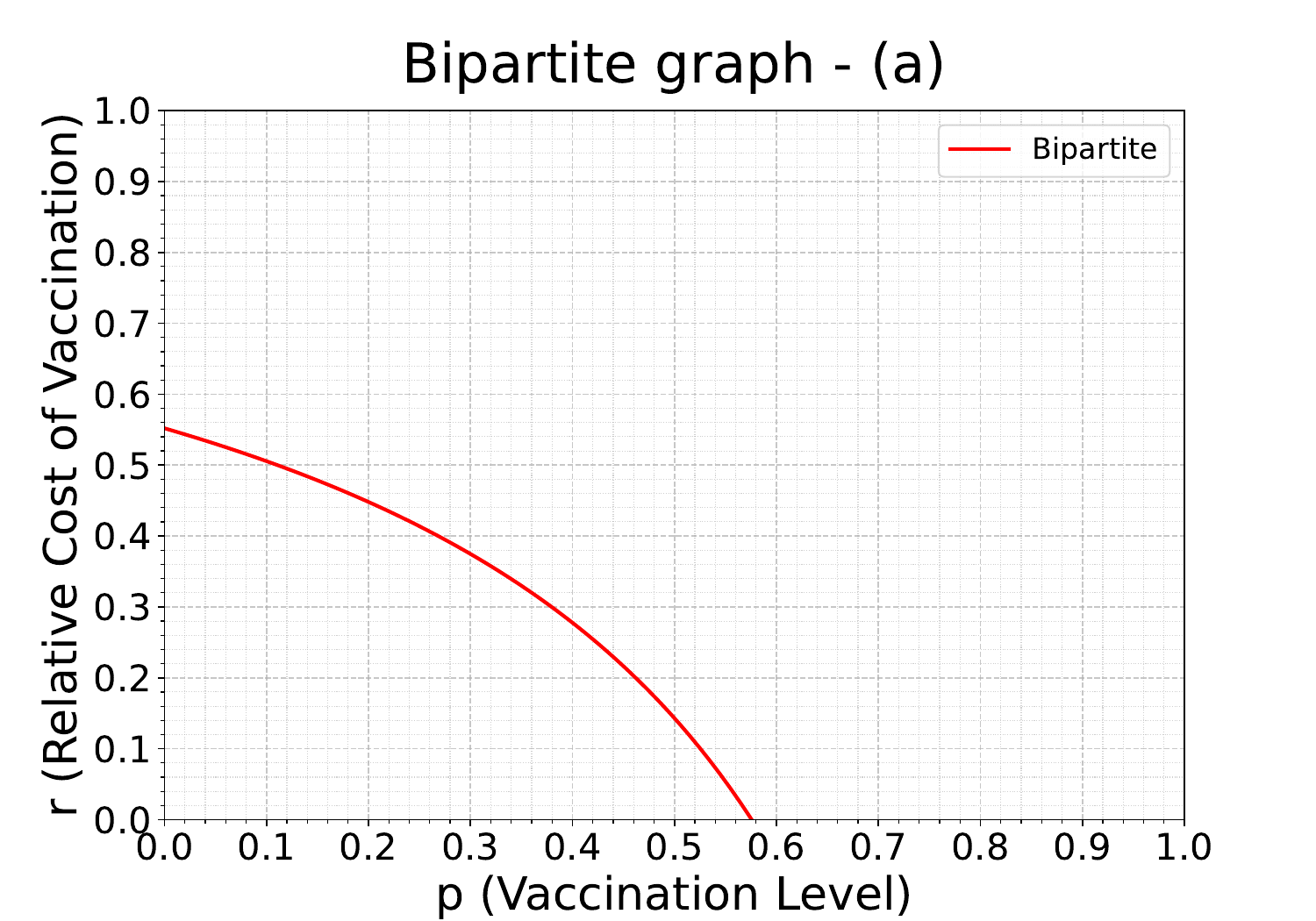}
    \end{subfigure}\hfill
    \begin{subfigure}{0.32\textwidth}
       \includegraphics[width=\textwidth]{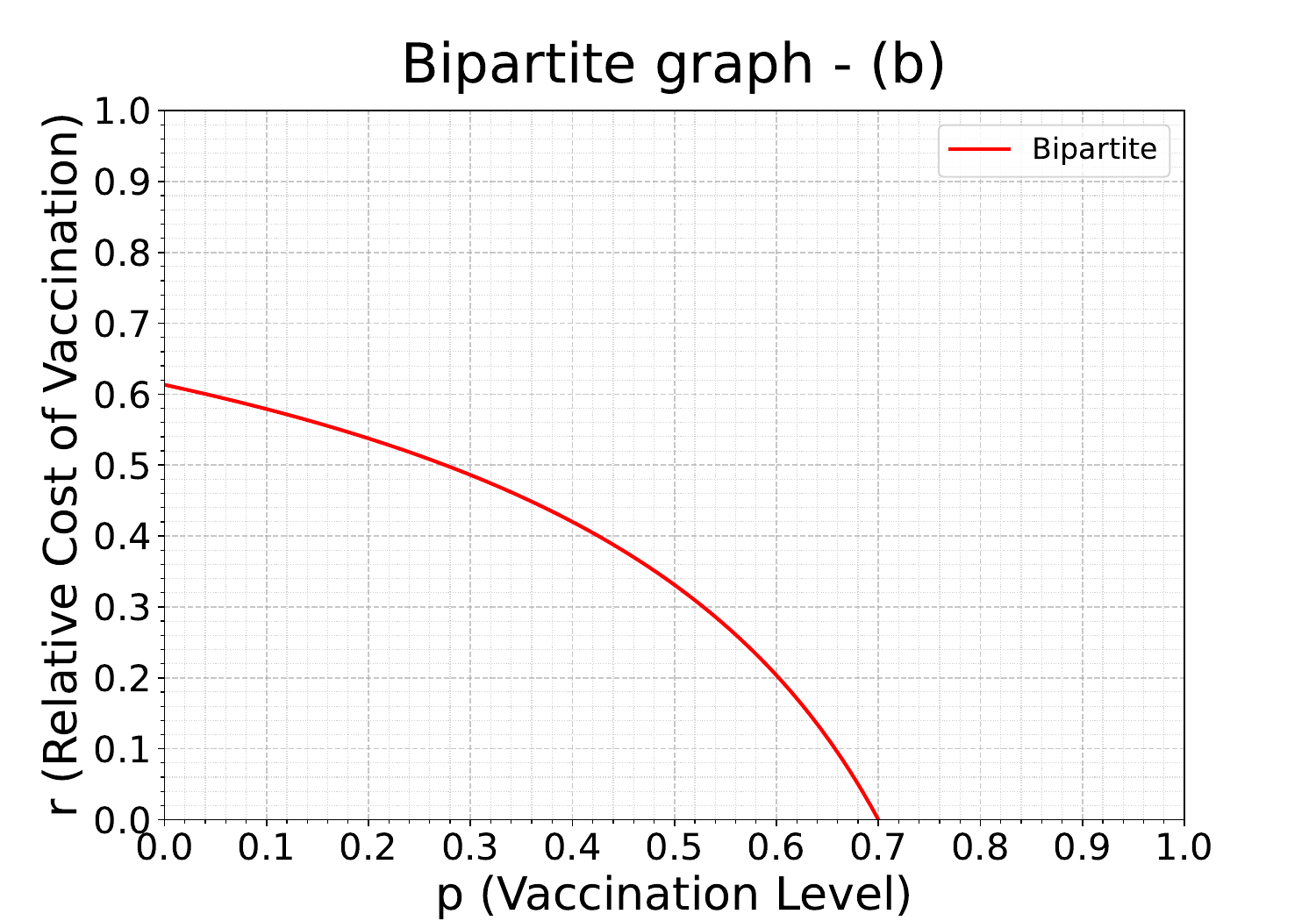}
    \end{subfigure}\hfill
    \begin{subfigure}{0.32\textwidth}
       \includegraphics[width=\textwidth]{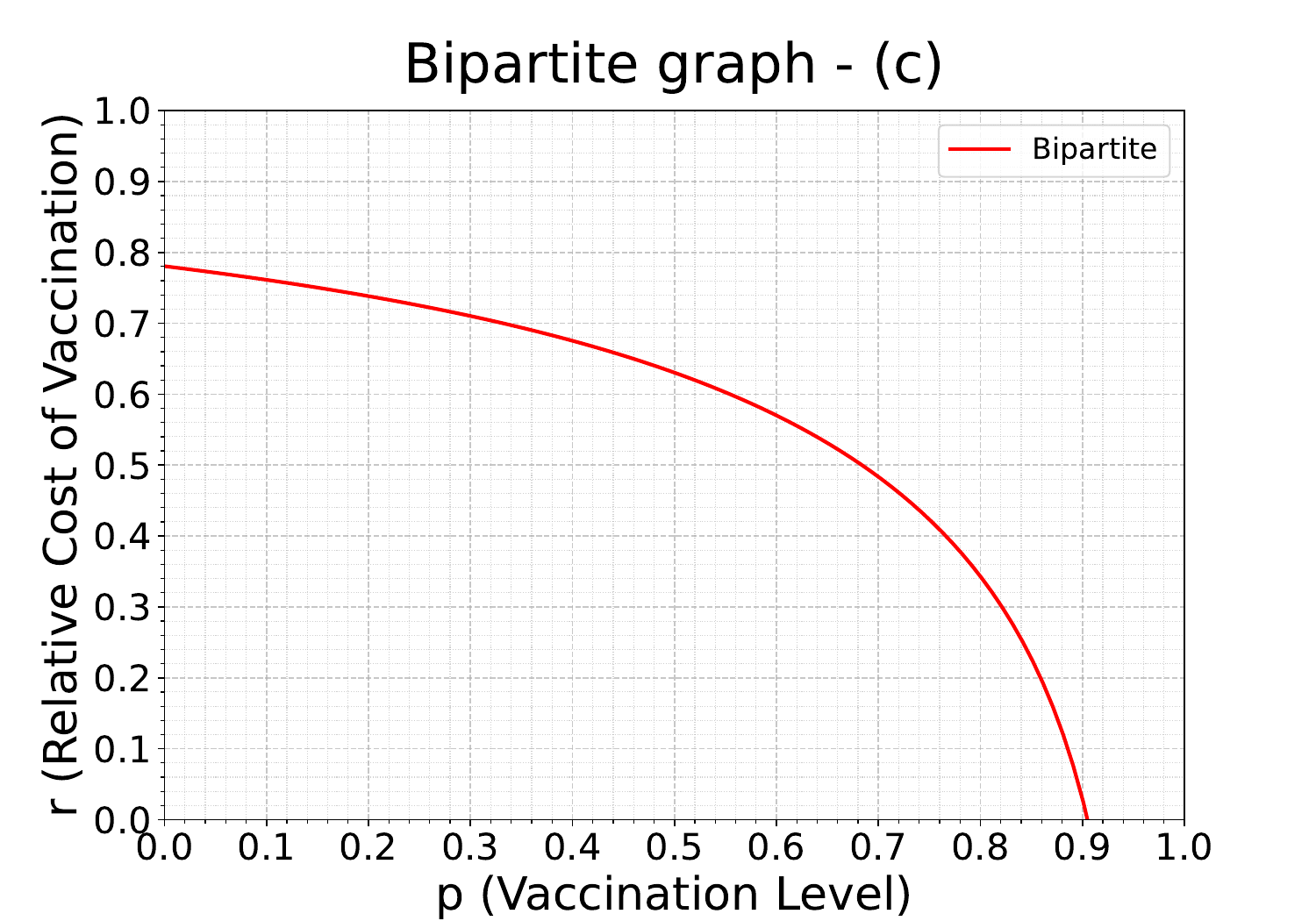}
    \end{subfigure}
    \caption{The first row shows a regular graph of degree $5$ (Regular graph - (a)), degree $10$ (Regular graph - (b)), and degree $20$ (Regular graph - (c)). The second row shows the bipartite graph: (a) $d_1 = 5$, $d_2 = 10$, (b): $d_1 = 5$, $d_2 = 20$, and (c): $d_1 = 100$, $d_2 = 10$.}
    \label{fig:degree_statistics}
    \end{figure}

In Figure~\ref{fig:degree_statistics} we compare regular and bipartite graphs corresponding to equation~\eqref{eq:r_complete} and equation~\eqref{eq:r_bipartite}, respectively. We use the following values for the numerical computation: initial infection $X_i = 0.4$, $b= 0.5$, $c=1/3$.
Figure~\ref{fig:degree_statistics} presents the solution of equation~\eqref{eq:r_complete} for the regular graph case (Regular graph - (a), Regular graph - (b), and Regular graph - (c) with degree $d = 5$, $d = 10$, and $d = 20$) and the bipartite graph case from equation~\eqref{eq:r_bipartite} for different degree $d_1$ and $d_1$ combinations. If $d_1$ = $d_2$, then equation~\eqref{eq:r_bipartite} become exactly the same as equation~\eqref{eq:r_complete}.

\section{Example networks}

In this section we present the data used in our study and the numerical results obtained from equations~\eqref{eq:infec_final} and~\eqref{eq:requation}, using the adjacency matrices of different networks to compare vaccination level required to achieve the evolutionarily stable vaccination level given the relative cost of vaccination.

\subsection{Data}
The numerical analysis in this study used publicly available real-world face-to-face contact networks collected by the SocioPatterns collaboration (\href{http://www.sociopatterns.org/}{sociopatterns.org}) in different real-world settings such as the Malawi village network \cite{ozella2021using} which consist of $86$ nodes and $355$ edges, primary school networks day $1$ (with $236$ nodes and $5899$ edges) and day $2$ (consisting of $238$ nodes and $5539$ edges) \cite{stehle2011high, gemmetto2014mitigation}, and the SFHH conference (with $403$ nodes and  $9565$ edges) \cite{genois2018can}. We also used the \textit{Facebook page networks} available \href{https://snap.stanford.edu/data/gemsec-Facebook.html}{snap.stanford.edu}, which represent mutual like relationships between verified Facebook pages such as Government and artists networks \cite{rozemberczki2019gemsec}. In these networks, each node corresponds to a Facebook page, while an edge between two nodes represents a mutual like interaction between the pages. The Government network consists of $7,057$ edges and $89,455$ edges, while the artist network consists of $50,515$ and	$819,306$ edges. Another network that we used is the Facebook network also from the Stanford Network Analysis Project (SNAP) (\href{https://snap.stanford.edu/data/egonets-Facebook.html}{snap.stanford.edu}), which consists of 4039 nodes and 88,234 edges described in \cite{leskovec2012learning}. We selected the above set of networks to provide a range of networks with different properties to explore the types of outcome that our model would predict.

We compute the degree variance to understand how much the degrees of the nodes deviate from the average degree.
The degree variance was calculated in the following manner: 
\begin{equation}
\text{Var}(k) = \frac{1}{N} \sum_{i=1}^{N} (k_i - \bar{k})^2
\label{eq:variance_k}
\end{equation}
where $k_i$ is the degree of node  $i$ (number of connections of node  $i$), $\bar{k}$ is the mean degree of the network, and $N$ is the total number of nodes in the network. 
The degree variance is an absolute measure, therefore, it is sensitive to the scale of the data. For instance, large networks or networks with larger mean degrees will have higher variances, even if their relative variability is similar. To account for this when quantifying the degree variability, we used the coefficient of variation $(CV(k)$ defined as the ratio of the standard deviation $\sigma_{k} = \sqrt{\text{Var}(k)}$ to the mean $\bar{k}$:
\begin{equation}
CV(k) = \frac{\sigma_{k}}{\bar{k}}
\end{equation}

The networks degree statistics (average degree $\bar{k}$, average degree variance $(Var(k))$, and the degree coeffient of variation ($CV(k)$)) are shown in Table~\ref{tab:degree_stats}.

\begin{table}[h!]
\centering
\begin{tabular}{|l|>{\centering\arraybackslash}p{4.8cm}|>{\centering\arraybackslash}p{3.7cm}|>{\centering\arraybackslash}p{1.5cm}|}
\hline
\textbf{Network} & \textbf{Degree variance ${Var(k)}$} & \textbf{Average degree} $\bar{k}$ & \textbf{CV($k$)} \\
\hline
Malawi Village         & 25.54    & 8.24   & 0.61 \\
Primary School Day 1   & 357.53   & 49.99  & 0.38 \\
Primary School Day 2   & 394.03   & 46.55  & 0.43 \\
SFHH Conference        & 908.63   & 47.47  & 0.64 \\
Facebook               & 2747.24  & 43.69  & 1.20 \\
Government             & 1373.58  & 25.35  & 1.46 \\
Artist                 & 4028.94  & 32.43  & 1.96 \\
\hline
\end{tabular}
\caption{Degree statistics for different networks used in this study.}
\label{tab:degree_stats}
\end{table}

\subsection{Numerical computation }

In this section, we present the results that come from solving equation \eqref{eq:infec_final} and equation 
\eqref{eq:requation} simultaneously and numerically using different network topologies. The goal here is, given any adjacency matrix, we should be able to work out what the appropriate evolutionarily stable vaccination level in the population is, for any given value of the relative cost of vaccination $r$. 

\begin{figure}[!ht]
\vspace{0.6cm}
  \centering
  \begin{minipage}[b]{0.47\linewidth} 
    \centering
    \begin{overpic}[width=\linewidth]{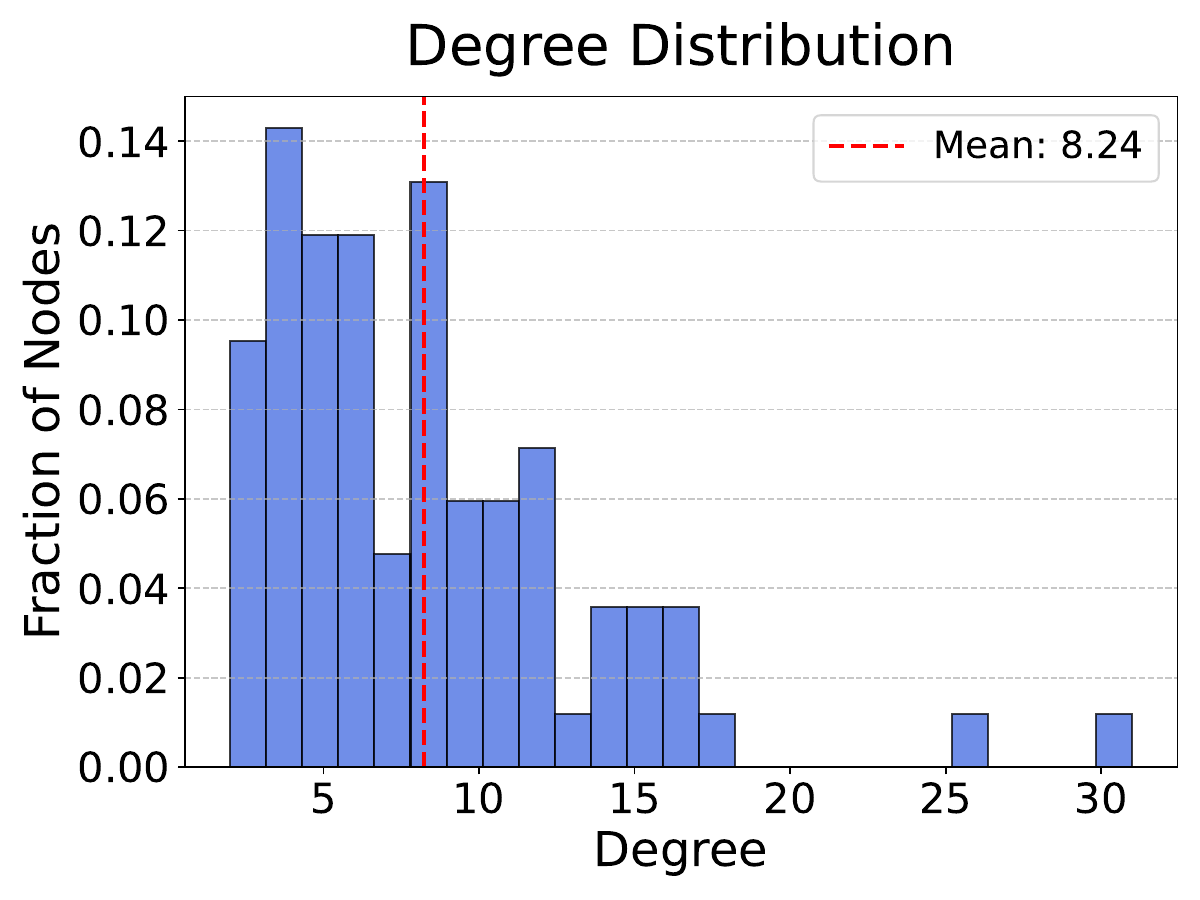}
      \put(7,75){\textbf{(a)}} 
    \end{overpic}
  \end{minipage}
  \hspace{0.02\linewidth} 
  \begin{minipage}[b]{0.49\linewidth}
    \centering
    \begin{overpic}[width=\linewidth]{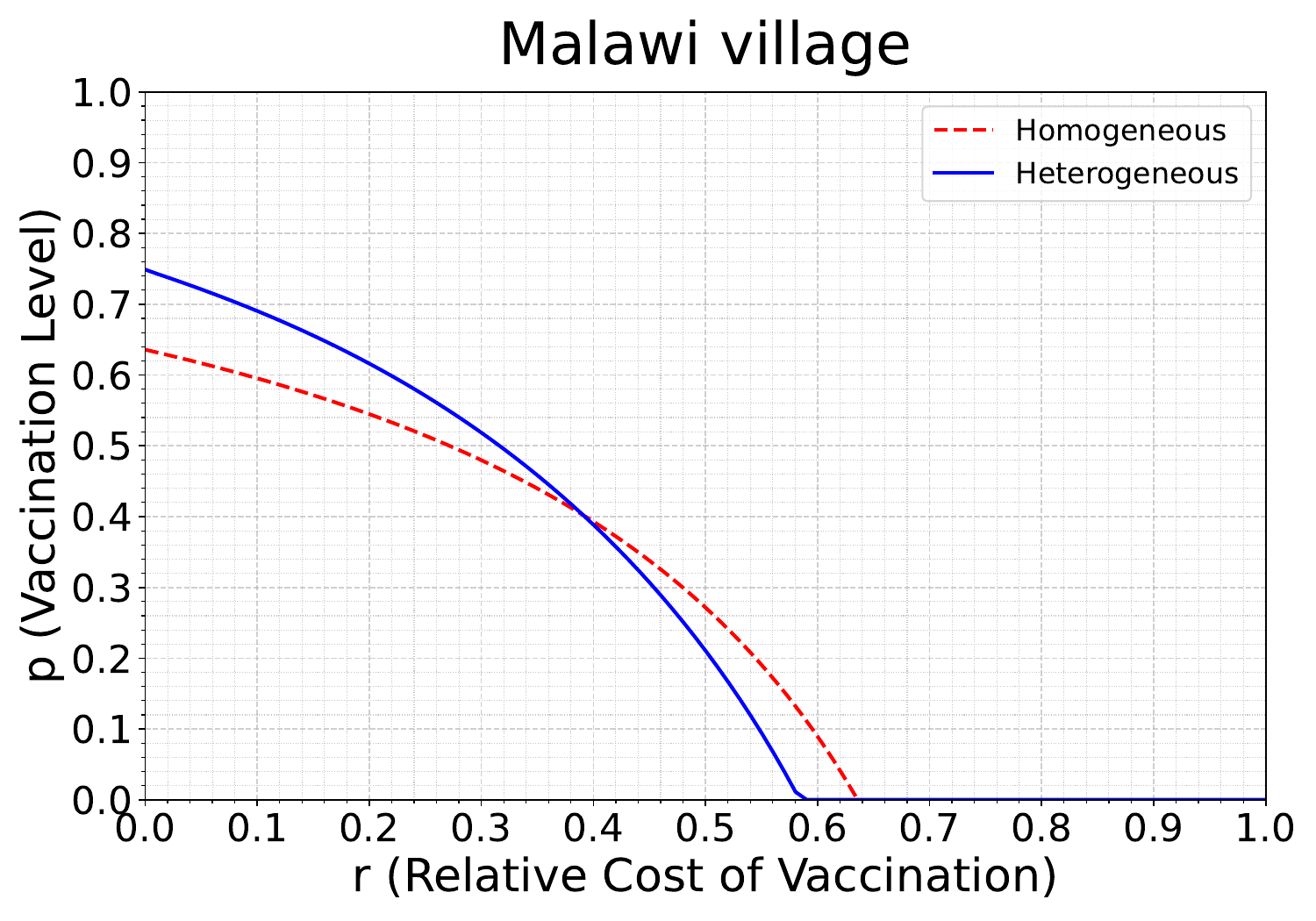}
      \put(7,71){\textbf{(b)}} 
    \end{overpic}
  \end{minipage}
  \caption{Analysis of the Malawi network: (a) Degree distribution of the Malawi village, (b) Vaccination level $p$ as a function of the relative cost of vaccination $r$. The red dashed line represents a homogeneous (well-mixed) population, while the blue solid line represents the real heterogeneous network.} 
  \label{fig:malawi}
\end{figure}

The average degree of the network of the Malawi village is $8.24$, meaning that, on average, each node (individual) in the network has about eight unique connections. The actual network of the Malawi village (represented by the blue curve) requires a higher vaccination coverage $p$ than its equivalent well-mixed population (represented by the red dashed line) of average degree $8$ for the relative cost of vaccination $r$ between $0$ and around $0.39$. However, beyond this point, the vaccination levels for the well-mixed equivalent of the Malawi village networks are higher than those for the actual network as shown in Figure~\ref{fig:malawi} (b). In the heterogeneous case of the Malawi village network nobody gets vaccinated as soon as the relative cost of vaccinations $r$ reaches $0.59$, while the well mixed case needed $r \approx 0.64$ for everybody to stop getting vaccinated.

\begin{figure}[!ht]
\vspace{0.6cm}
  \centering
  \begin{minipage}[b]{0.46\linewidth}
    \centering
    \begin{overpic}[width=\linewidth]{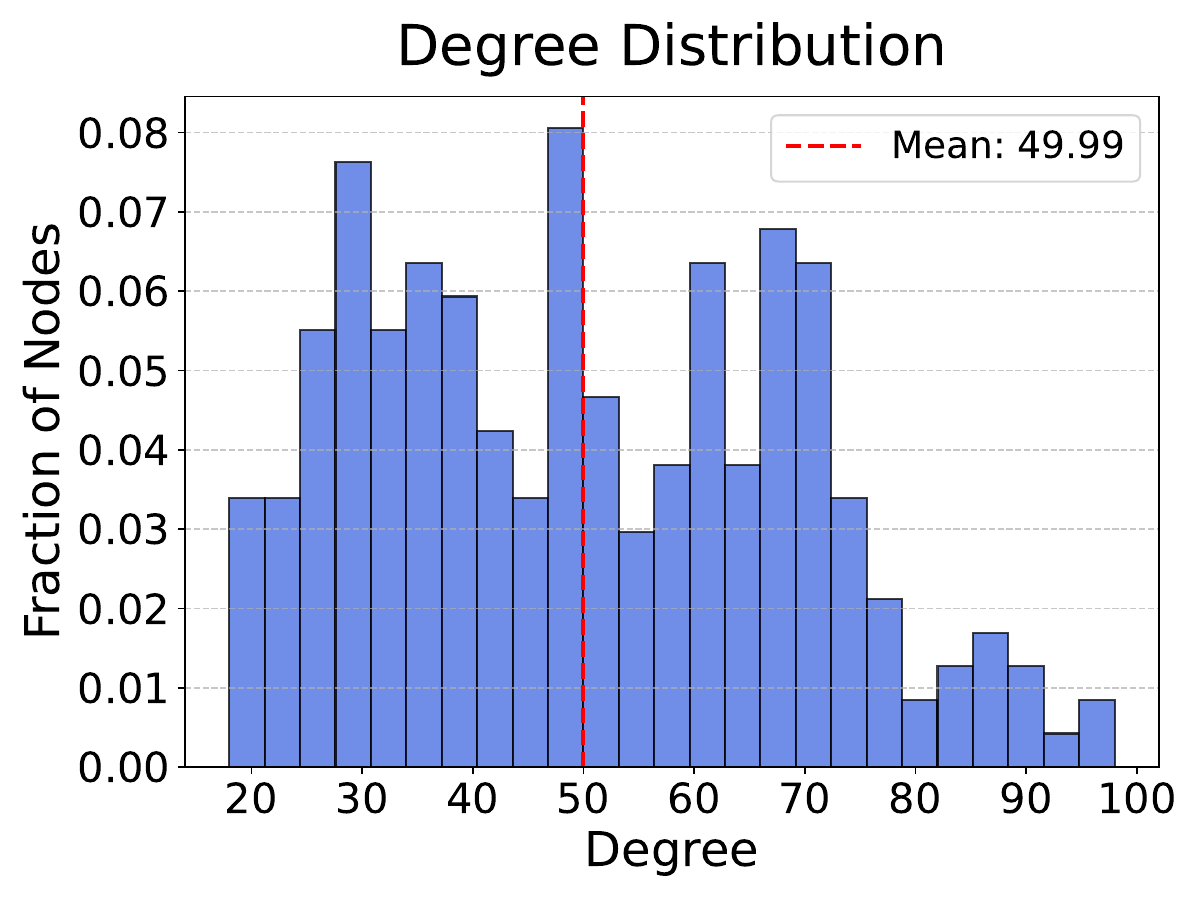}
      \put(7,75){\textbf{(a)}} 
    \end{overpic}
  \end{minipage}
  \hspace{0.02\linewidth} 
  \begin{minipage}[b]{0.49\linewidth}
    \centering
    \begin{overpic}[width=\linewidth]{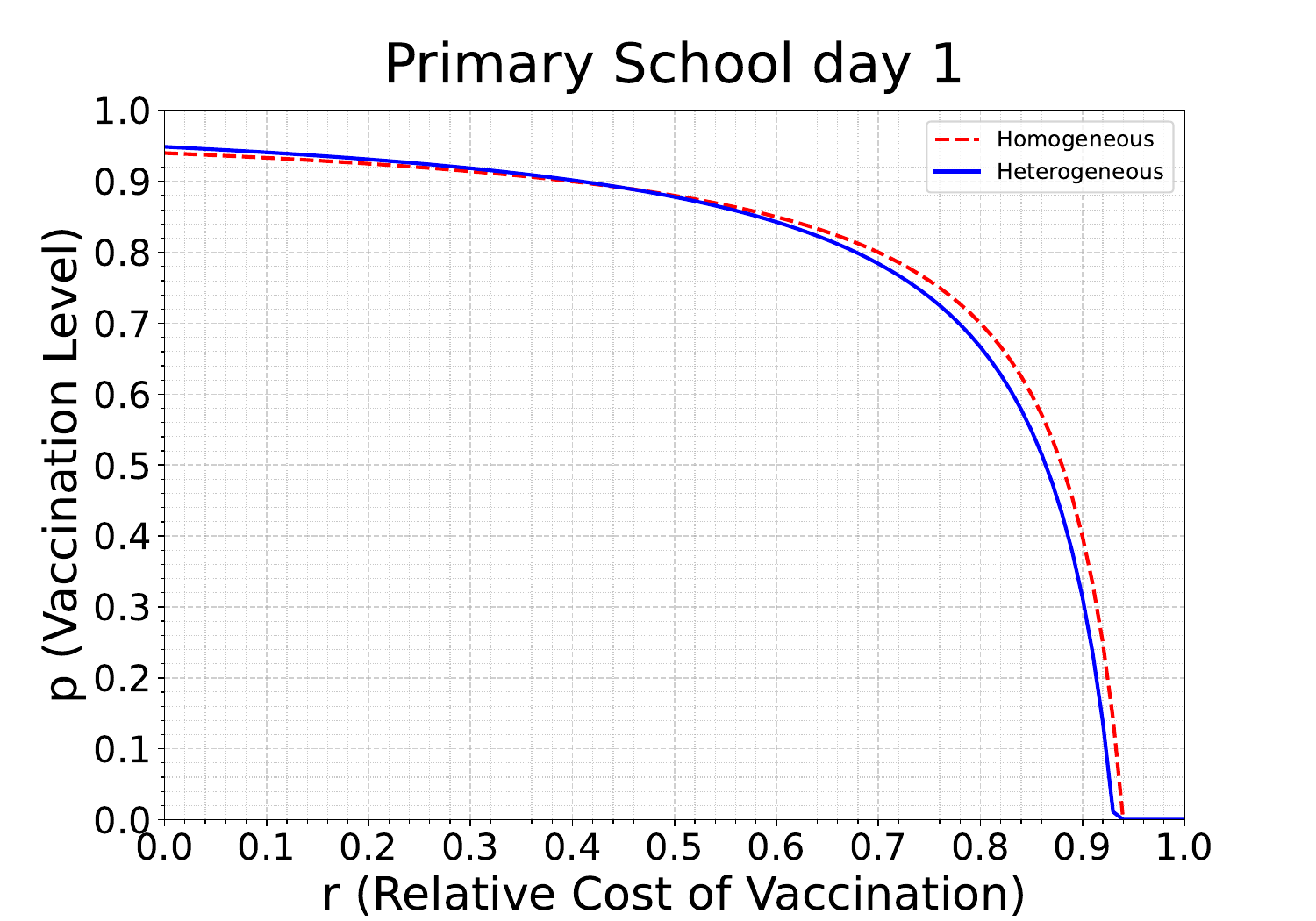}
      \put(7,71){\textbf{(b)}} 
    \end{overpic}
  \end{minipage}
  \caption{Analysis of Primary School Day 1 data. (a) Degree distribution of the Primary School Day 1 network, (b) Vaccination level $p$ as a function of the relative cost of vaccination $r$. The red dashed line represents a homogeneous (well-mixed) population, while the blue solid line represents the real heterogeneous network.}
  \label{fig:primary day 1}
\end{figure}

\begin{figure}[!ht]
\vspace{0.6cm}
  \centering
  \begin{minipage}[b]{0.47\linewidth} 
    \centering
    \begin{overpic}[width=\linewidth]{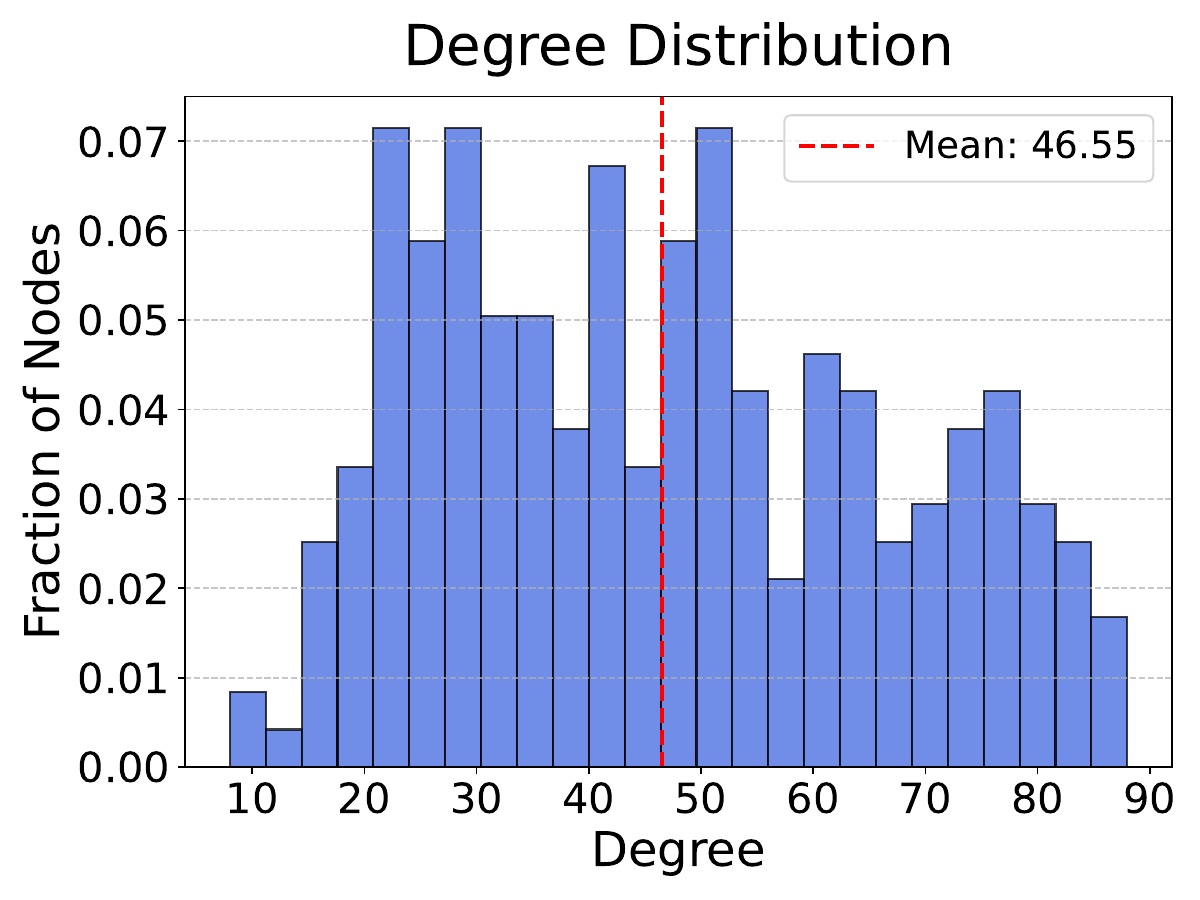}
      \put(7,75){\textbf{(a)}} 
    \end{overpic}
  \end{minipage}
  \hspace{0.02\linewidth} 
  \begin{minipage}[b]{0.49\linewidth}
    \centering
    \begin{overpic}[width=\linewidth]{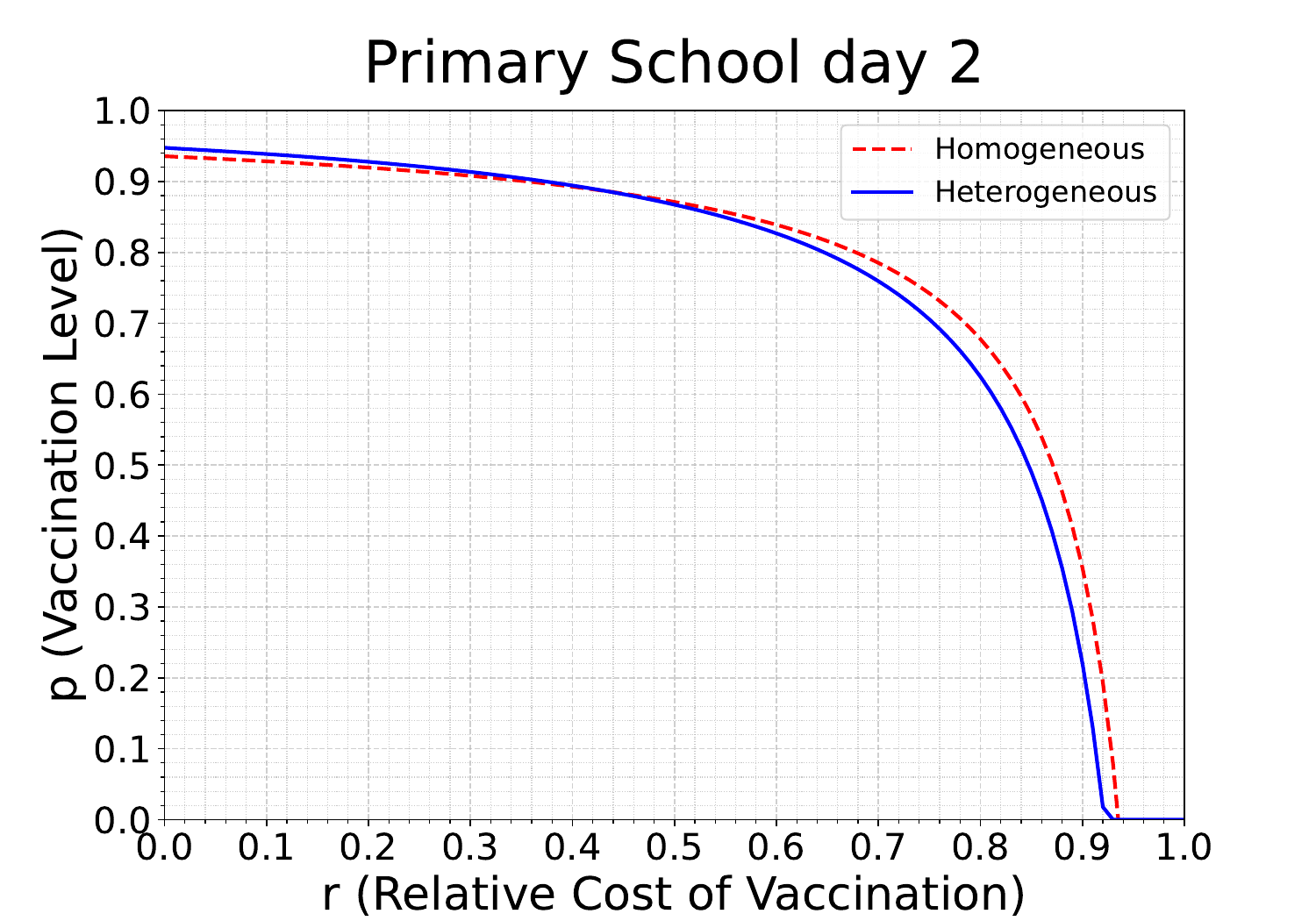}
      \put(7,73){\textbf{(b)}} 
    \end{overpic}
  \end{minipage}
  \caption{Analysis of Primary School Day 2 data: (a) Degree distribution of the Primary School Day 2 network, (b) Vaccination level $p$ as a function of the relative cost of vaccination $r$. The red dashed line represents a homogeneous (well-mixed) population, while the blue solid line represents the real heterogeneous network.}
  \label{fig:primary day 2}
\end{figure}

The degree distribution plots in Figure \ref{fig:primary day 1} (a) and  Figure \ref{fig:primary day 2} (a) show the number of connections each student has, with a higher mean degree on Day 1 (49.99) compared to Day 2 (46.55), that is students had more interactions on the first day. 

The vaccination coverage plots in (b) of both Figure \ref{fig:primary day 1} and Figure \ref{fig:primary day 2} show that as the cost of vaccination increases, the proportion of vaccinated individuals decreases in both homogeneous (well-mixed) and heterogeneous networks. However, for $r$ between $0$ and $0.5$, the evolutionarily stable vaccination levels for the heterogeneous case are slightly higher than those required for the homogeneous case (for both days of the primary school network). The vaccination levels for the well-mixed population is higher than that of the heterogeneous population for $r \ge 0.5$ on both day 1 and day 2 of the primary school networks. The primary schools networks (Day 1 and Day 2) also maintain high levels of vaccination even for a high relative cost of vaccination and it takes a high relative cost of vaccination ($0.93$ and $0.94$) for the vaccination levels to drop to zero in both primary school networks. Furthermore, the vaccination levels for the homogeneous and the heterogeneous cases of the primary school networks are almost the same.

\begin{figure}[!ht]
\vspace{0.6cm}
  \centering
  \begin{minipage}[b]{0.46\linewidth} 
    \centering
    \begin{overpic}[width=\linewidth]{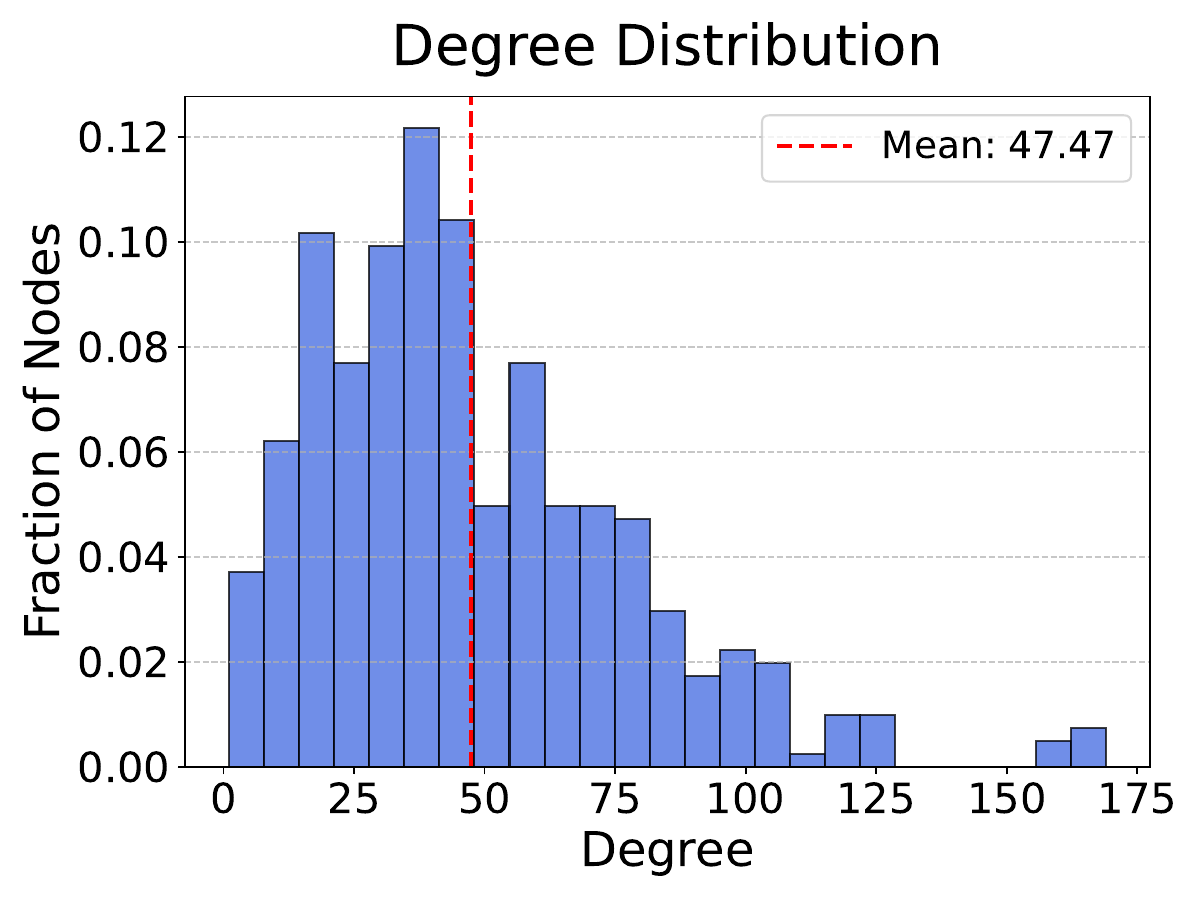}
      \put(7,75){\textbf{(a)}} 
    \end{overpic}
  \end{minipage}
  \hspace{0.02\linewidth} 
  \begin{minipage}[b]{0.49\linewidth}
    \centering
    \begin{overpic}[width=\linewidth]{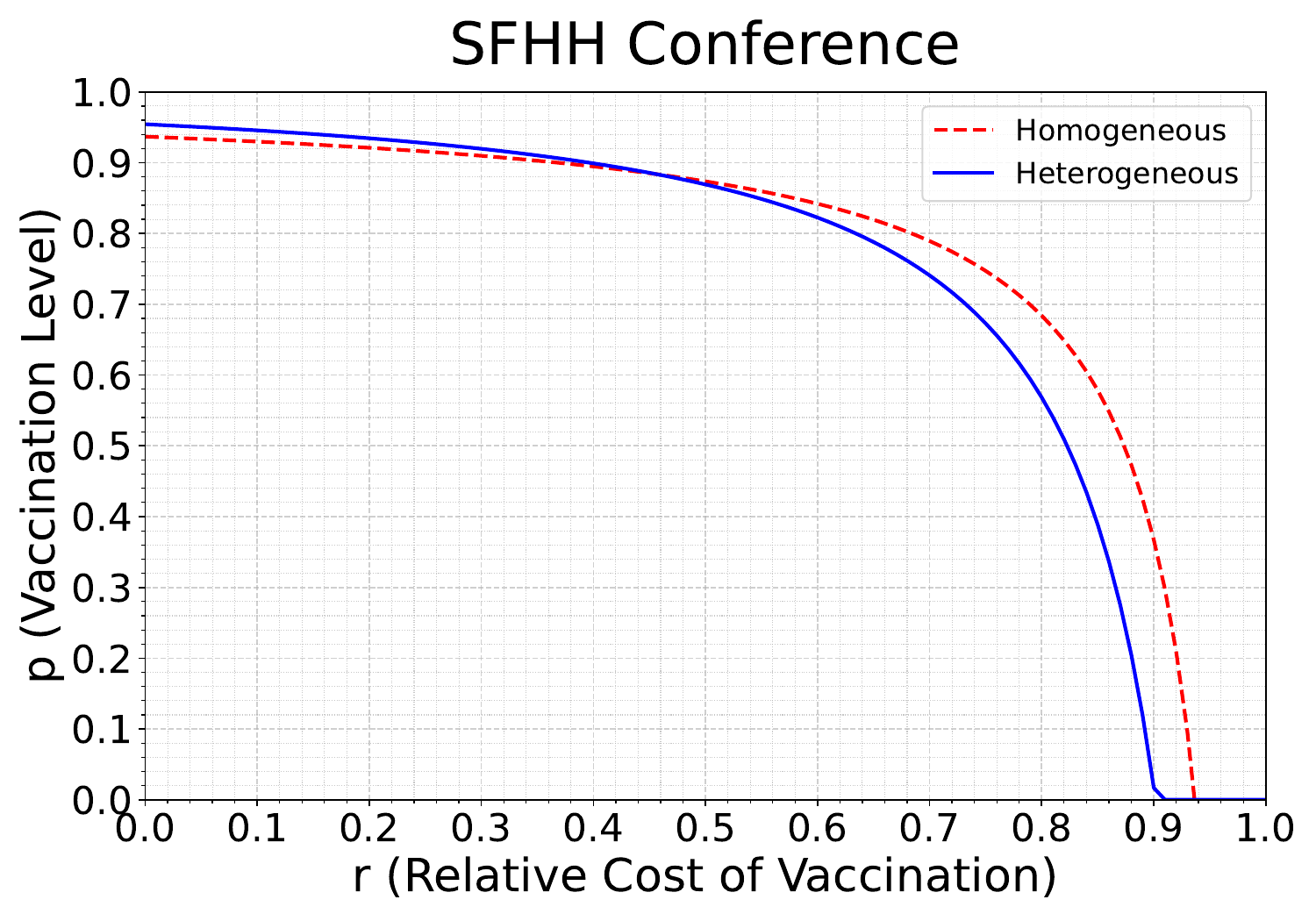}
      \put(7,72){\textbf{(b)}} 
    \end{overpic}
  \end{minipage}
  \caption{Analysis of SFHH conference dataset: (a) Degree distribution of the SFHH conference network, (b) Vaccination level $p$ as a function of the relative cost of vaccination $r$. The red dashed line represents a homogeneous (well-mixed) population, while the blue solid line represents the real heterogeneous network.}
  \label{fig:SFHH}
\end{figure}

Figure \ref{fig:SFHH} panel (a) shows the degree distribution of the SFHH conference network (with average degree of  47.47), where most nodes have a relatively low degree and a few nodes have higher degrees. This degree distribution is different from the primary networks we saw in Figure \ref{fig:primary day 1} and Figure \ref{fig:primary day 2}, but the vaccination levels for the well-mixed population become higher than that of the heterogeneous network at almost the same values of $r$ (around $r \ge 0.48$).  The SFHH conference follow a similar trend to those observed in the primary school networks (Day 1 and Day 2), however the differences at high $r$ is bigger than that for the primary school networks.

\begin{figure}[!ht]
\vspace{0.6cm}
  \centering
  \begin{minipage}[b]{0.46\linewidth} 
    \centering
    \begin{overpic}[width=\linewidth]{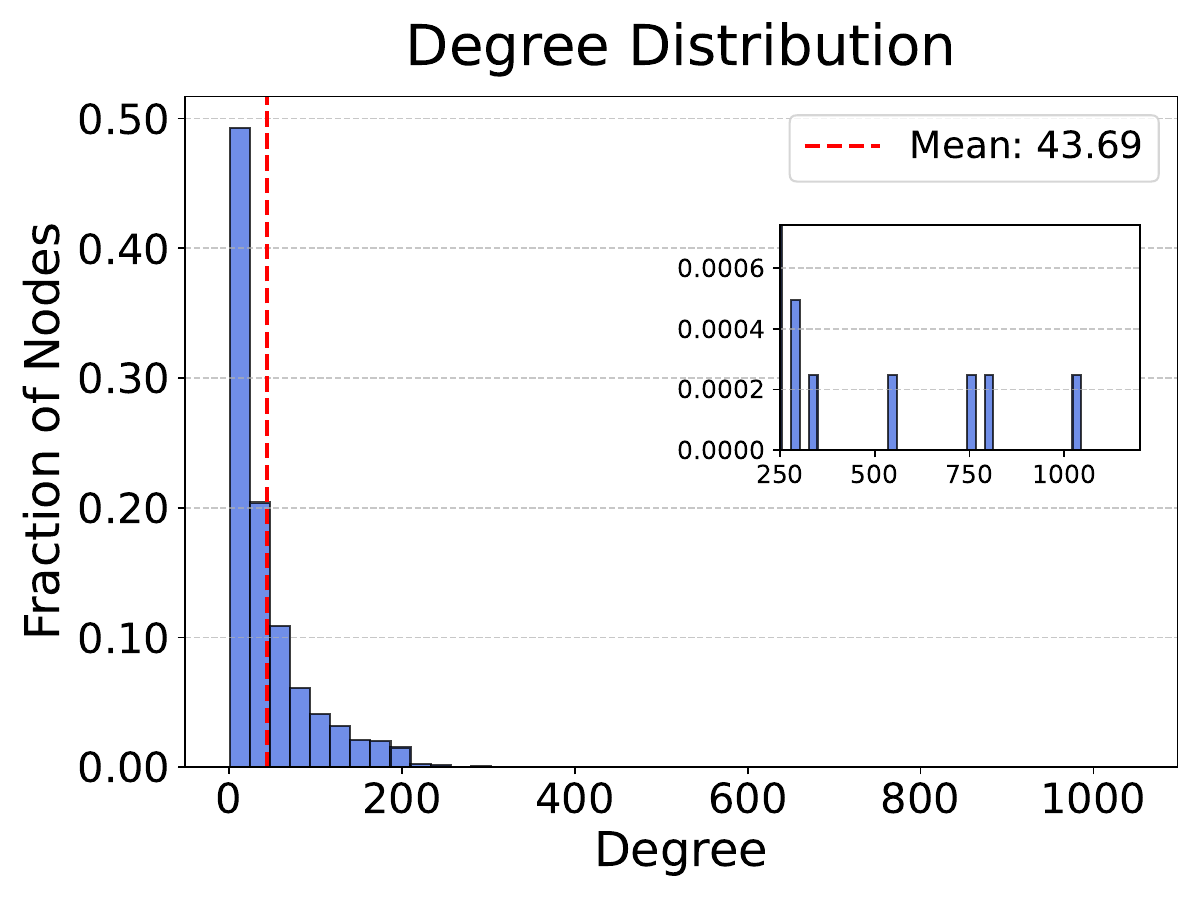}
      \put(7,75){\textbf{(a)}} 
    \end{overpic}
  \end{minipage}
  \hspace{0.02\linewidth} 
  \begin{minipage}[b]{0.49\linewidth}
    \centering
    \begin{overpic}[width=\linewidth]{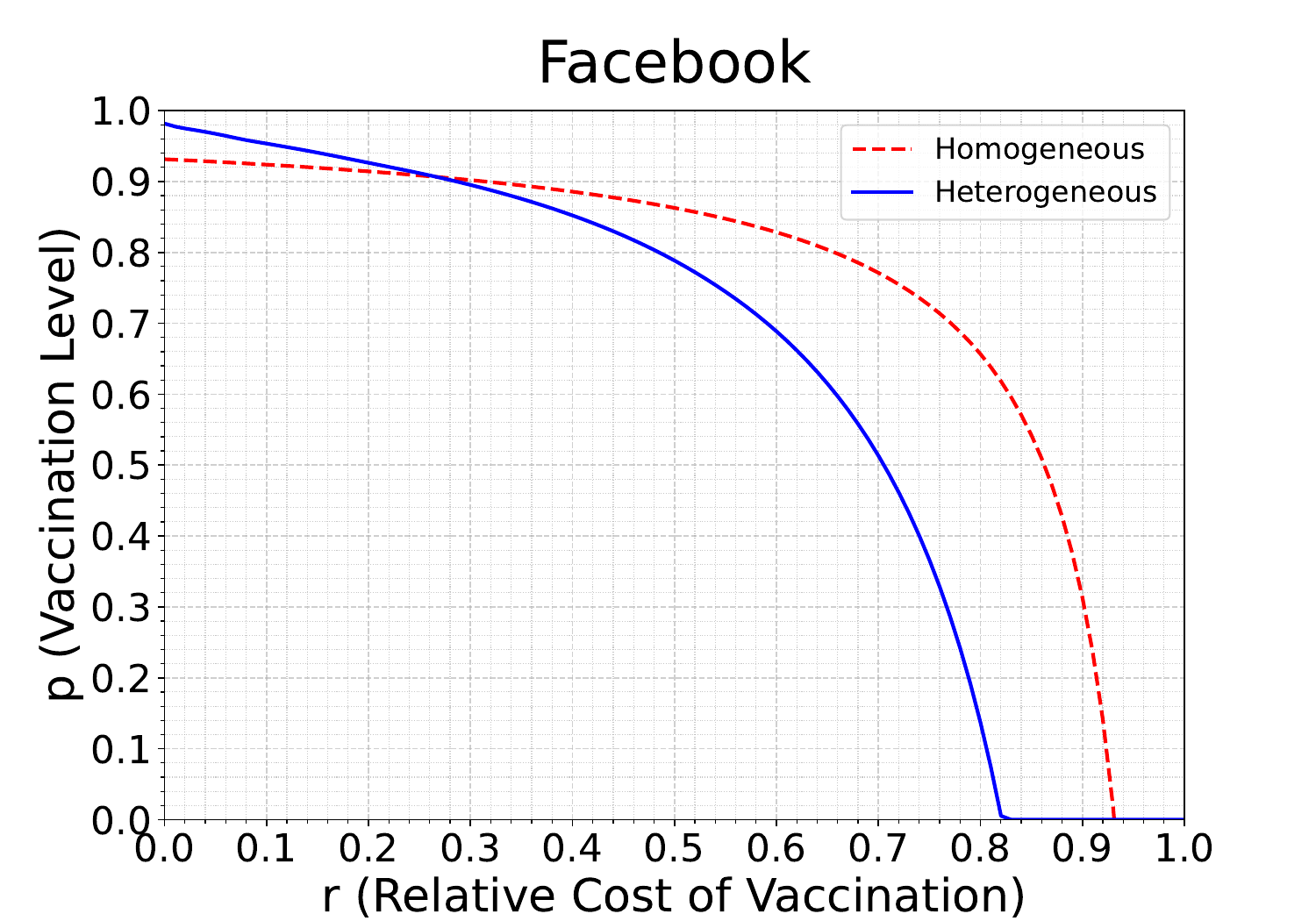}
      \put(7,70){\textbf{(b)}} 
    \end{overpic}
  \end{minipage}
  \caption{Facebook network analysis: (a) degree distribution and (b) vaccination level $p$ as a function of the relative cost of vaccination $r$. The red dashed line represents the homogeneous case (well-mixed population), while the blue solid line represents the real heterogeneous network.}
  \label{fig:facebook}
\end{figure}

In Figure \ref{fig:facebook}, both the homogeneous (red dashed line) and heterogeneous (blue solid line) cases show a decline but a much sharper decline for the actual Facebook network (blue line) vaccination level $p$ as the relative cost of vaccination r increases. The well-mixed equivalent of the Facebook network require higher vaccination levels for the relative cost of vaccination $r$ ($\ge 0.3$). This indicates that as vaccination becomes more costly, fewer individuals opt to vaccinate.

The government and artist networks have average degrees of 25.35 and 32.43, respectively (both have fewer connections per node compared to the Facebook network). Their results follow a similar trend as observed in the Facebook networks, but the vaccine level is lower than the one for the Facebook network. For these 3 networks (Facebook, Government and Artist), as the relative cost of the vaccine $r$ increases the difference between the vaccine level for the actual network (blue line) and its homogeneous equivalent (red line) becomes wider and clearer. The homogeneous equivalent of these networks have higher evolutionarily stable vaccination levels as the relative cost of vaccination increases.
\begin{figure}[!ht]
\vspace{0.6cm}
  \centering
  \begin{minipage}[b]{0.46\linewidth} 
    \centering
    \begin{overpic}[width=\linewidth]{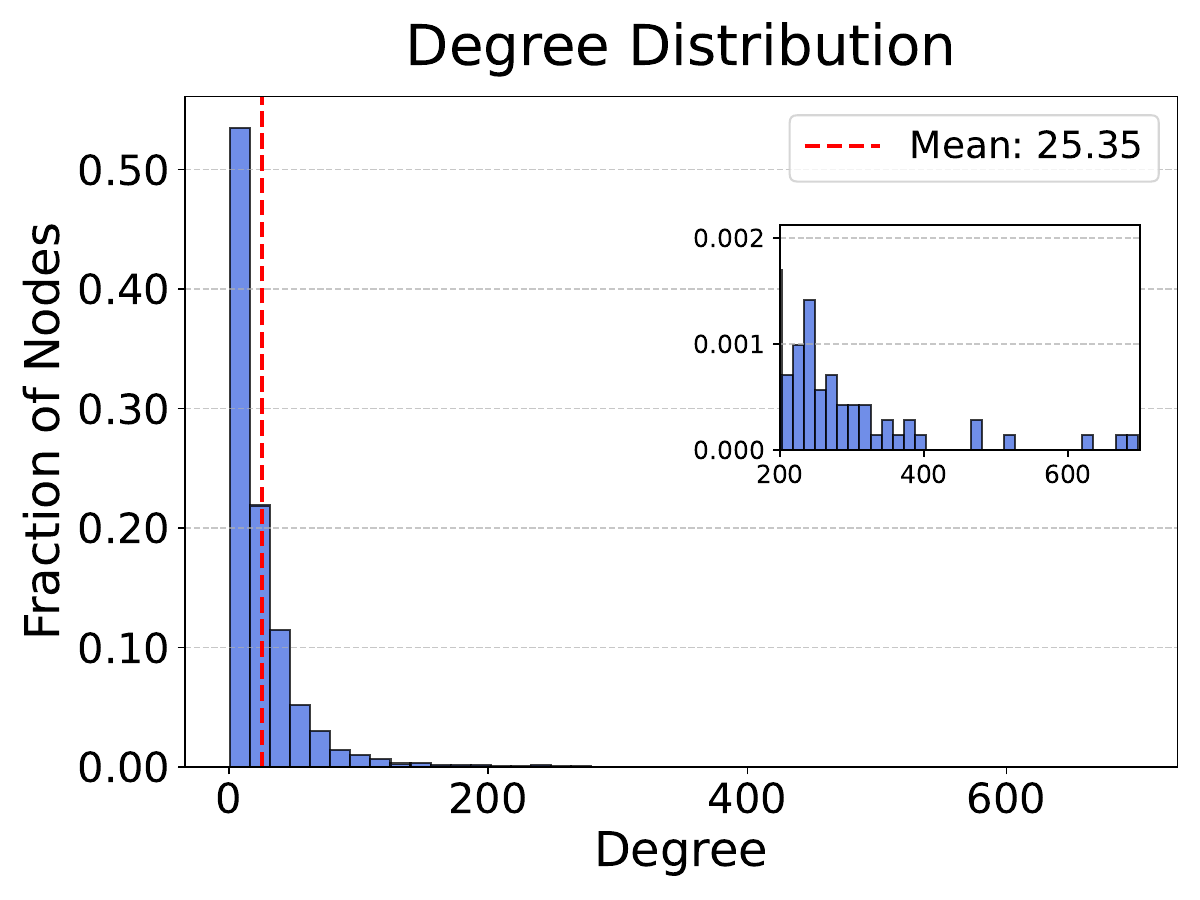}
      \put(7,75){\textbf{(a)}} 
    \end{overpic}
  \end{minipage}
  \hspace{0.02\linewidth} 
  \begin{minipage}[b]{0.49\linewidth}
    \centering
    \begin{overpic}[width=\linewidth]{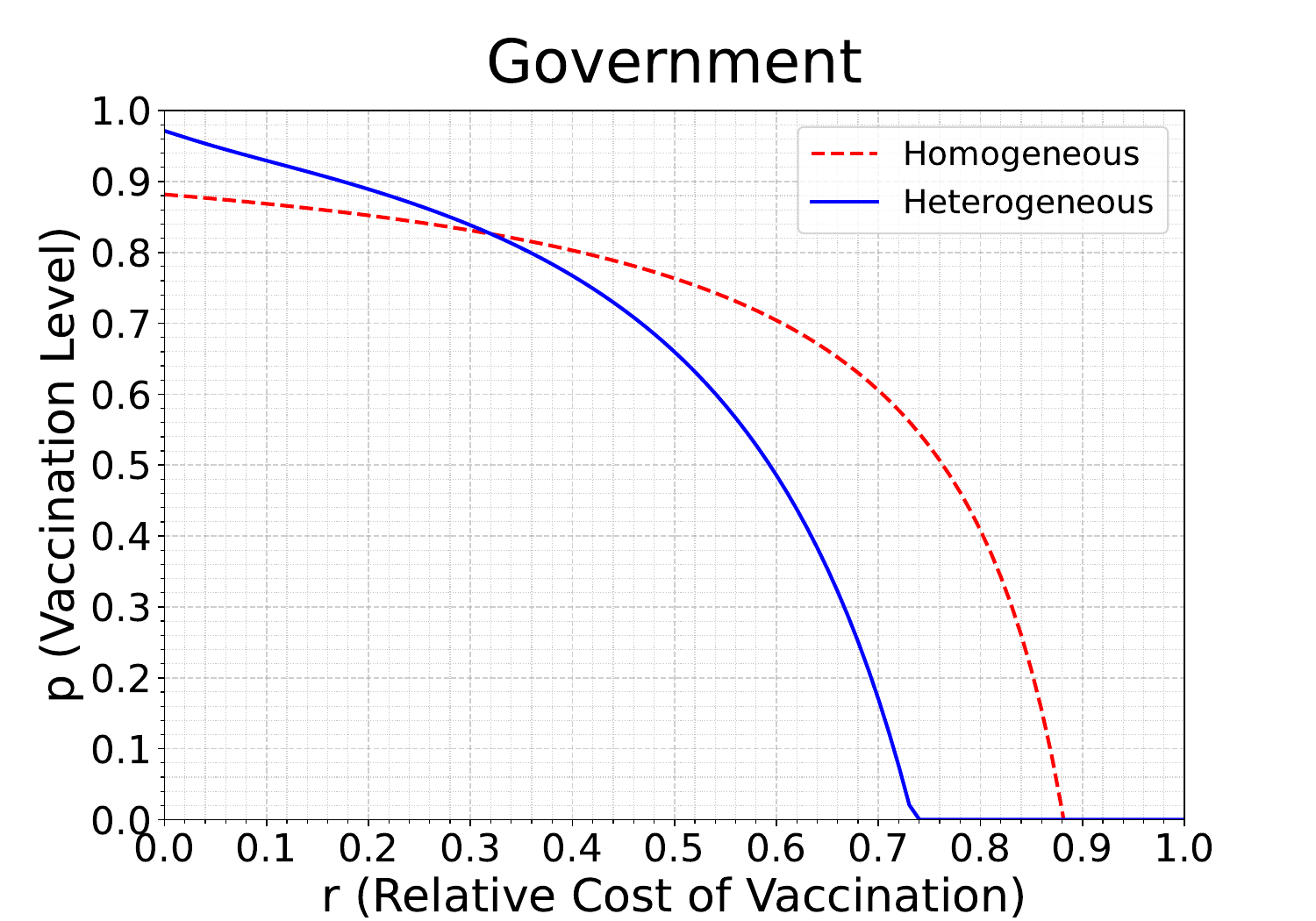}
      \put(7,70){\textbf{(b)}} 
    \end{overpic}
  \end{minipage}
  \caption{Government network analysis: (a) Degree distribution of the Government network and (b) Vaccination level $p$ as a function of the relative cost of vaccination $r$. The red dashed line represents a homogeneous (well-mixed) population, while the blue solid line represents the real heterogeneous network.}
  \label{fig:government}
\end{figure}

\begin{figure}[!ht]
\vspace{0.6cm}
  \centering
  \begin{minipage}[b]{0.46\linewidth} 
    \centering
    \begin{overpic}[width=\linewidth]{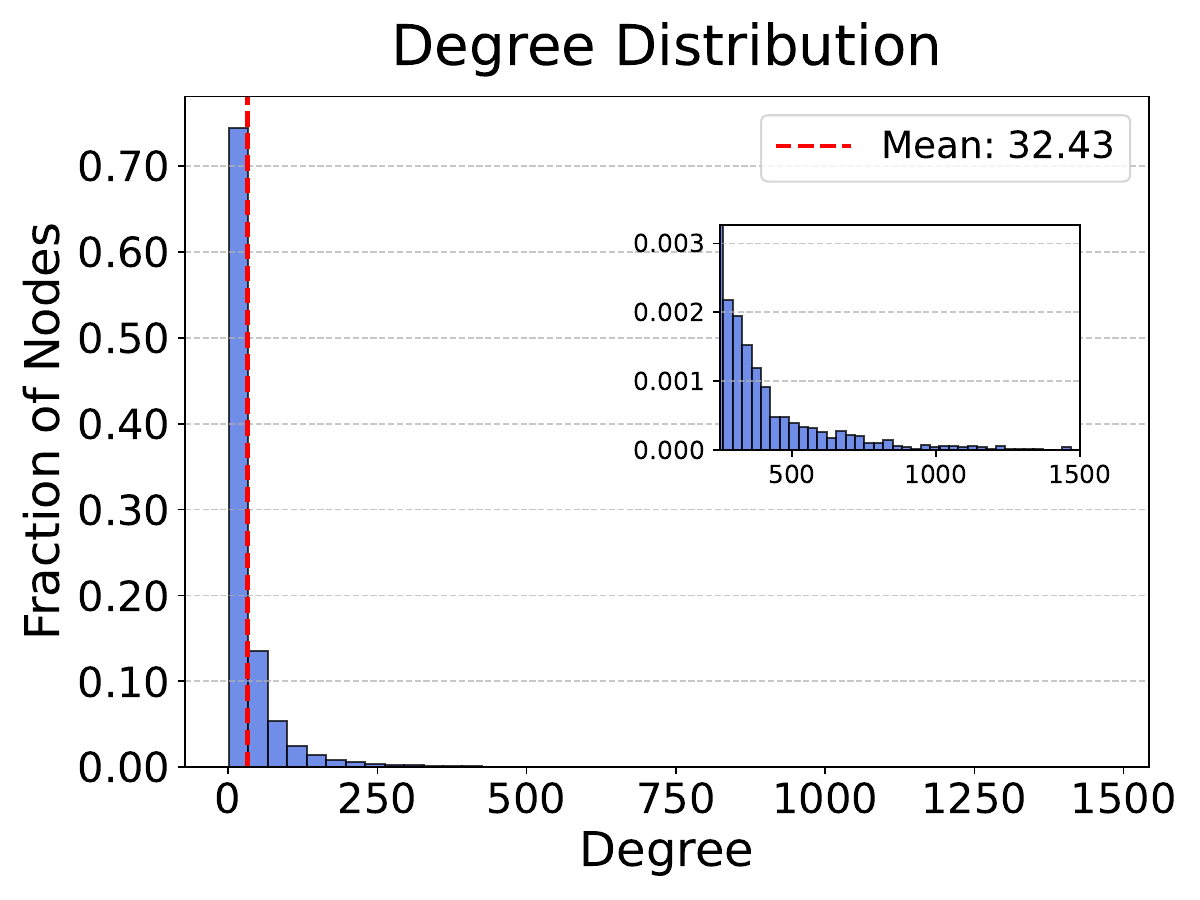}
      \put(7,75){\textbf{(a)}} 
    \end{overpic}
  \end{minipage}
  \hspace{0.02\linewidth} 
  \begin{minipage}[b]{0.49\linewidth}
    \centering
    \begin{overpic}[width=\linewidth]{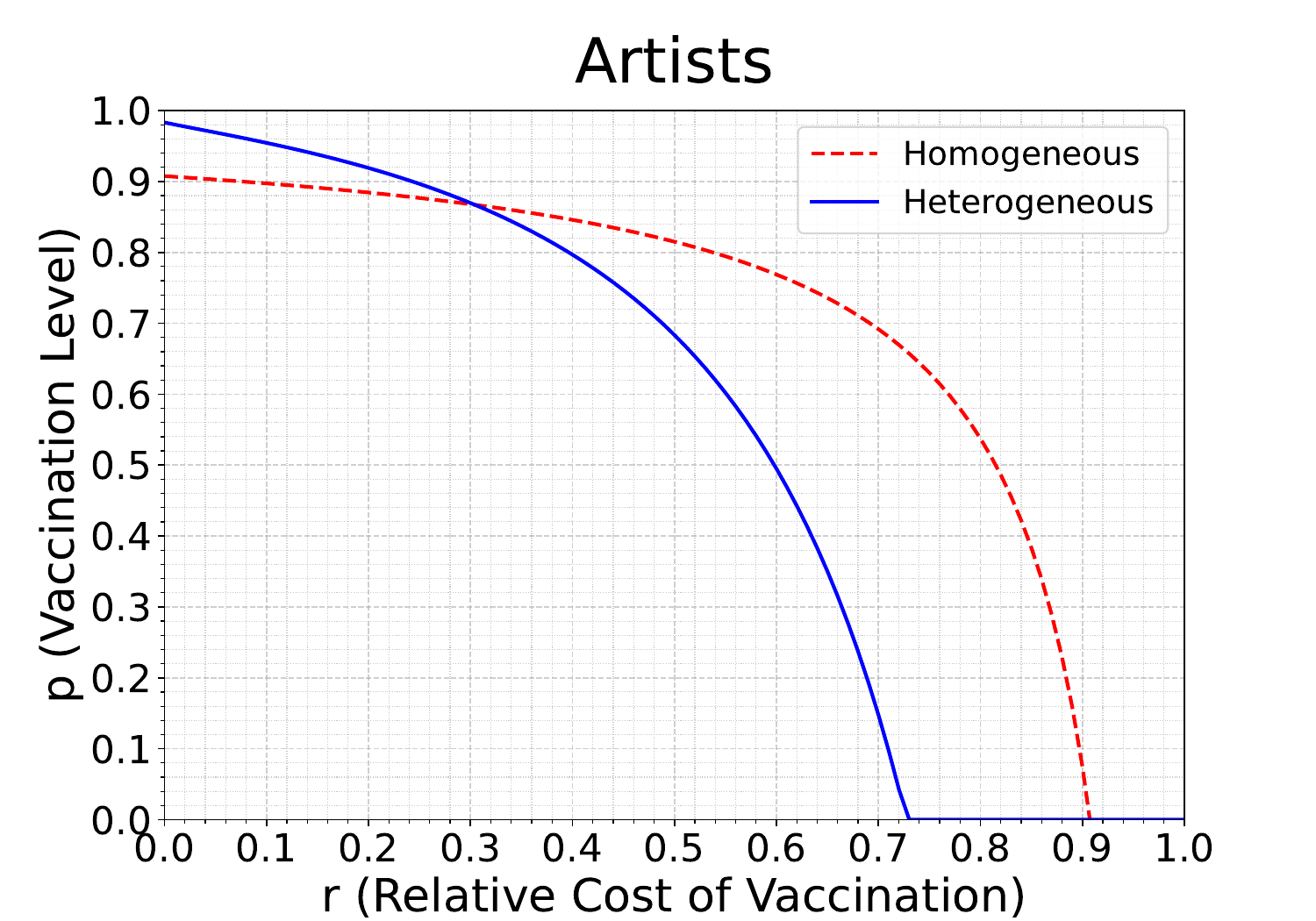}
      \put(7,70){\textbf{(b)}} 
    \end{overpic}
  \end{minipage}
  \caption{Artist network analysis: (a) Degree distribution of the Artists network and (b) Vaccination level p as a function of the relative cost of vaccination r. The red dashed line corresponds to a homogeneous (well-mixed) population, while the blue solid line represents the real heterogeneous network.}
  \label{fig:artist}
\end{figure}

\newpage

\section{Discussion}

Our results show that the vaccination coverage (fraction of individuals vaccinated) that will emerge varies across different network topologies. As the relative cost of vaccination $(r)$ increases, vaccination uptake decreases in all networks (individuals become less willing to get vaccinated when vaccines are expensive relative to infection). This finding supports the idea that when the relative cost of vaccination is high, meaning that the perceived risk of vaccination (vaccine morbidity risk) outweighs the perceived risk of infection (infection morbidity risk), then there is no incentive to vaccinate (the dominant strategy is to never vaccinate) as also found in Bauch and Earn \cite{bauch2004vaccination}. This is because individuals perceive a higher risk from the vaccine compared to the disease, leading to a pure non-vaccinator strategy. Across all networks, the vaccination plots show a consistent trend: heterogeneous networks generally require higher or similar vaccine coverage $p$ than well-mixed populations at low values of the relative cost of vaccination $r$, with the magnitude of the gap reflecting the degree of heterogeneity and connectivity. Another important common trend for all the networks is that, as the relative cost of vaccination $r$ increases, the difference between the vaccination levels for the heterogeneous and homogeneous becomes more pronounced, with the homogeneous having higher levels than the heterogeneous case. In fact, this is consistent with the results of Elbasha and Gumel \cite{elbasha2021vaccination}, who used a non-game theoretical model to consider herd immunity thresholds for different population structures. They showed that when a population is heterogeneous, the minimum herd immunity threshold can be substantially lower than that of the homogeneous case. The networks with low degree variability, particularly the primary school (Day 1 and Day 2) networks, maintain higher vaccination coverage for high cost, and the evolutionarily stable vaccination level is high for most values of $r$ compared to networks with high degree variability such as Facebook, Government, and Artist networks. 
In networks with low degree variability, the vaccination levels, given the relative cost of vaccinations, are almost equal to their homogeneous equivalent networks (given by equation \eqref{eq:r_complete}). The SFHH conference network shows a similar pattern to the primary school day $1$ and day $2$ networks. However, the difference in vaccination levels between actual networks and the homogeneous equivalent is bigger in the SFHH conference network compared to the primary school networks. This is because the SFHH conference network has a slightly higher degree of variability than the primary school networks. On the other hand, the more heterogeneous networks, such as Facebook, Government, and Artist, have a high degree of variability and require very high vaccination coverage at a low relative cost of vaccination compared to other networks (Malawi village, Primary school, and SFHH conference). This is consistent with results from \cite{fu2011imitation}, which suggest that network heterogeneity promotes vaccination in networks whose degree distribution follows a power law at a lower relative cost of vaccination. In these networks, the differences between the vaccination levels for the heterogeneous case and the homogeneous equivalent is bigger from $r \approx 0.3$. Another important observation is that, in the more heterogeneous networks (Facebook, Government and Artist), the vaccination levels drop to zero at a lower relative cost of vaccination compared to the primary schools and SFHH conference networks.

The predictions from our work depend on our model assumptions, and in particular, the key assumption of a uniform vaccination level across the network. Whilst we believe this is often a reasonable assumption, when individuals have experience of general infection levels but do not make sophisticated locality-based decisions, it is possible that some members of the population who consider themselves at higher or lower risk due to their connectivity would choose different vaccination decisions. This would lead to a more complicated modelling scenario, and we leave this case for later work.

\section*{Funding}

This project has received funding from the European Union’s Horizon 2020 Research and Innovation Programme under Grant Agreement No. 955708.

\printbibliography
\end{document}